\documentclass[12pt]{article}
\usepackage[tmargin=2cm,lmargin=2cm,rmargin=2cm,bmargin=2cm]{geometry}
\usepackage[utf8]{inputenc}
\usepackage{setspace}
\usepackage{parskip}
\usepackage{amssymb}
\usepackage{amsmath}
\usepackage{graphicx}
\usepackage[english]{babel}
\usepackage{fancyhdr}
\usepackage[T1]{fontenc}
\usepackage{gensymb}
\usepackage{epstopdf}
\usepackage{siunitx}
\usepackage{graphicx} 
\usepackage{float}
\usepackage{cite}
\usepackage{wasysym}
\usepackage{mathtools}
\usepackage{authblk}
\usepackage{titlesec}
\usepackage[numbers,square,super]{natbib}

\titlelabel{\thetitle.\quad}
\makeatletter
\newcommand\bigDiamond{\mathop{\mathpalette\bigDi@mond\relax}}
\newcommand\bigDi@mond[2]{%
  \vcenter{\hbox{\m@th
    \scalebox{\ifx#1\displaystyle 2\else1.2\fi}{$#1\Diamond$}%
  }}%
}
\newcommand\bigLozenge{\mathop{\mathpalette\bigL@zenge\relax}}
\newcommand\bigL@zenge[2]{%
  \vcenter{\hbox{\m@th
    \scalebox{\ifx#1\displaystyle 2\else1.2\fi}{$#1\blacklozenge$}%
  }}%
}
\makeatother

\begin{document}
\Large
\title{\textbf{Plasmonically Enhanced Reflectance of Heat Radiation from Low-Bandgap Semiconductor Microinclusions}}
\large
\author[1*]{Janika Tang} 
\author[1*]{Vaibhav Thakore} 
 \author[1,2,3]{Tapio Ala-Nissila}
\affil[1]{COMP CoE at the Department of Applied Physics, Aalto University School of Science, FIN-00076 Aalto, Espoo, Finland}
\affil[2]{Department of Physics, Brown University, Providence, Rhode Island 02912-1843, USA}
\affil[3]{Department of Mathematical Sciences and Department of Physics, Loughborough University, Loughborough LE11 3TU, UK}
\maketitle
\text{*Email: vthakore@knights.ucf.edu} (Corresponding author)\\
\text{*Email: janika.tang@aalto.fi}  (Corresponding author)\\
\normalsize 
\setcounter{secnumdepth}{0}
\doublespacing
\section{Abstract}
Increased reflectance from the inclusion of highly scattering particles at low volume fractions in an insulating dielectric offers a promising way to reduce radiative thermal losses at high temperatures. Here, we investigate plasmonic resonance driven enhanced scattering from microinclusions of low-bandgap semiconductors (InP, Si, Ge, PbS, InAs and Te) in an insulating composite to tailor its infrared reflectance for minimizing thermal losses from radiative transfer. To this end, we compute the spectral properties of the microcomposites using Monte Carlo modeling and compare them with results from Fresnel equations. The role of particle size-dependent Mie scattering and absorption efficiencies, and, scattering anisotropy are studied to identify the optimal microinclusion size and material parameters for maximizing the reflectance of the thermal radiation. For composites with Si and Ge microinclusions we obtain reflectance efficiencies of $57$ - $65 \%$ for the incident blackbody radiation from sources at temperatures in the range $400$ - $1600$ \degree C. Furthermore, we observe a broadbanding of the reflectance spectra from the plasmonic resonances due to charge carriers generated from defect states within the semiconductor bandgap. Our results thus open up the possibility of developing efficient high-temperature thermal insulators through use of the low-bandgap semiconductor microinclusions in insulating dielectrics. 

\textbf{Keywords:} radiative heat transport, high-temperature insulators, localized surface plasmon resonance, Mie scattering and infrared reflectance

\setcounter{secnumdepth}{3}
\section{Introduction}
Efficient thermal insulation at a given temperature must reduce unwanted heat exchange with the surrounding environment that occurs primarily through the twin modes of conductive and radiative heat transfer. Designing an efficient thermal insulator thus involves a subtle tradeoff between minimizing conductive heat loss by optimizing the porosity of an insulating material, \textit{e.g.} with microstructured air-pockets, and simultaneously ensuring that there is no significant thermal loss through increased radiative heat transfer \cite{fricke,xie, zeng, zhang,milandri}. This approach works well for low temperature applications. However, under high temperature conditions radiative heat transfer becomes the dominant mode of thermal losses \cite{xie}. In such cases, decreasing the porosity of the material to prevent radiative losses becomes unfeasible as an alternative because it inevitably also leads to higher conductive losses. Therefore, a strategy for designing an efficient thermal insulator for high temperature applications must carefully balance the two phenomena. The ability to tailor the broadband infrared reflectance to minimize radiative losses has important implications for providing efficient thermal insulation under high temperature conditions and in applications such as furnaces, fire protection, gas-turbine engines, redirecting heat in photovoltaic systems, in energy-efficient buildings, etc. \cite{viskanta, berdahl, padture, wijewardane}. 

A vast amount of literature exists on new materials for coatings and paints doped with metal/metal-oxide pigments or dyes that is focused on obtaining increased absorbance or reflectance of solar radiation \cite{smith, slovick, Synnefa, levinson2005}. These coatings or paints are referred to as `cool' or `hot' depending on whether they enhance diffuse reflectance through scattering or enable spectrally selective absorption in the near-infrared wavelength (NIR) regime \cite{laaksonen2014, smith-deller,baneshi2009, levinson2007, johnson}. These materials, while excellent for facilitating effective harnessing of solar energy in photovoltaic devices or for thermal management in buildings and vehicles, are however not suitable for use as thermal insulators at high temperatures because of their high thermal conductivities \cite{slovick,laaksonen2013}. Multilayer dielectric materials used in thermal barrier coatings offer an alternative but are prohibitively expensive to fabricate and maintain for structurally complex systems \cite{fink, hughes}. In this regard, an attractive low-cost alternative is offered by thermal insulators such as aerogels that are characterized by remarkably low thermal conductivities. However, aerogels are almost transparent to the NIR wavelengths  (${3}$-$8$ $\mathrm{\mu m}$) rendering them unsuitable for use in high temperature environments \cite{xie}. Aerogel based thermal insulators therefore require the use of opacifiers for improving insulation at high temperatures wherein radiative transfer losses dominate \cite{xie}. Opacifiers are typically particles of refractory metal-oxides, carbides or nitrides that are randomly distributed at high mass fractions in aerogels to enable multiple scattering of thermal radiation and thereby improve diffuse reflectance \cite{xie, zeng,yu, zhao}. 

Recently, localized surface plasmon resonances (LSPRs) in randomly distributed metallic nanoparticles on surfaces and in films have been exploited to demonstrate controlled reflectance \cite{laaksonen2014,laaksonen2013,moreau}. LSPRs arise due to a confinement of the collective oscillations (plasmons) of free charge carriers on the surface of a micro or nanoparticle driven by the electromagnetic field of the incident radiation of wavelength greater than or comparable to the size of the particle \cite{petryayeva}. This results in enhanced scattering and absorption resonances that can be controlled with the geometry, size, dielectric environment and the spatial distribution of the particles  \cite{laaksonen2014,laaksonen2013,chen, temple, schelm2003, schelm2005, petryayeva, luther}. Although LSPRs in metallic particles can be tailored to modify reflectance, the tunability of their frequency response lies only in either the ultraviolet or visible spectrum of the electromagnetic radiation. Furthermore,  besides the regime of frequency response, the high thermal conductivity of metallic particles makes them unsuitable for use as opacifiers in insulators for high temperature applications. However, low-bandgap semiconductors characterized by relatively low-thermal conductivities exhibit LSPRs that can be excited by the incident heat radiation in the infrared regime \cite{luther}. Low-bandgap semiconducting inclusions therefore hold excellent promise as opacifiers in high temperature insulators. In this study, we focus our investigation on the effect of the plasmonic resonance induced enhanced scattering on the diffuse reflectance of thermal radiation from insulator dielectrics with low-bandgap semiconducting microinclusions.  

Radiative heat transport in materials can be modeled using several different methods that include numerical methods for solving the radiative transfer equation \cite{mishra}, ray-tracing based on geometrical optics \cite{melamed, simmons, stamnes}, flux based methods \cite{KM,niklasson,vargas1997,maheu1984} and Monte Carlo models \cite{mcml, zaccanti,briton, eddowes}. Numerical methods for solving the radiative transfer equation that employ a finite number of angular intensities such as the discrete transfer method (DTM), discrete ordinates method (DOM) and the finite volume method (FVM) typically require some kind of an assumption of angular isotropy for scattering \cite{mishra}. The radiation element method by the ray emission model ($\mathrm{REM}^2$) also employs a finite number of angular intensities but gets around this difficulty by considering scattering anisotropy through the use of a delta function approximation for the scattering phase function \cite{maruyama}. In general, these methods can be applied to complex geometries but they also tend to limit radiation transport to certain discrete directions thereby affecting their accuracy. The flux-based methods employ coupled ordinary differential equations to model radiative transport in two-dimensional media along the normal direction \cite{KM,niklasson,vargas1997,maheu1984}. The two-flux Kubelka-Munk (KM) \cite{KM} and the extended KM radiative transfer models \cite{vargas1997}, frequently employed due to their ease of implementation, are some of the oldest flux-based methods available for diffuse and collimated incident radiation respectively. However, the KM methods are applicable only to optically thick films with non-absorbing particles or to films with highly scattering and weakly absorbing particles with size-parameters larger than the Rayleigh limit \cite{vargas1997}. Improvements upon the KM models account for backward and forward fluxes of diffuse and collimated radiation separately through the incorporation of additional flux channels \cite{vargas1997}. The most widely used of these methods is the generalized four-flux model due to Vargas and Niklasson \cite{niklasson, vargas1997} based on the four-flux model proposed by Maheu \textit{et al.} \cite{maheu1984}. However, in the case of media characterized by large anisotropic scattering the generalized four-flux method requires an evaluation of the average path-length parameters using the extended Hartels theory \cite{vargas2}. On the other hand, Monte Carlo methods based on tracking packets of incident radiation (henceforth referred to as photons) in two or three dimensions are highly accurate and applicable to anisotropic media with multiple scattering without requiring the evaluation of any average path-length parameters or the use of a finite number of angular intensities \cite{maheu}. Thus, here we use a Monte Carlo method in conjunction with Mie theory for modeling radiation transport in a microcomposite dielectric insulator with spherical semiconducting microinclusions at low volume fractions.

Recently, Slovick \textit{et al.} have experimentally demonstrated the tailoring of the diffuse infrared reflectance of up to $90\%$ for LPC paints with microscale inclusions of single-crystal hexagonal Boron Nitride platelets (h-BN) albeit at an unusually high h-BN volume fraction of $f=0.5$ \cite{slovick}. Gonome \textit{et al.} have also demonstrated up to $90\%$ near-infrared broadband reflectances for cool coatings with submicron copper-oxide (CuO) particles at low volume fractions ranging from $f=0.02$ to $0.05$ \cite{gonome}. However, these high reflectances were obtained for coatings on highly reflecting white substrates while coatings on black substrates yielded significantly lower reflectances of about $35-40\%$ \cite{gonome}. Also, currently there exist no studies that systematically investigate the effect of Mie parameters for microparticles on maximizing the reflectance of incident thermal radiation from composites or coatings. Thus, the key objective of our study is to understand the role of the particle size-dependent Mie scattering $Q_{\mathrm{sca}}$ and absorption $Q_{\mathrm{abs}}$ efficiencies and the scattering anisotropy $g$ in designing insulating composites with low-bandgap semiconductor microinclusions at low volume fractions $f$ to maximize the reflectance of the incident thermal radiation. To this end, we compute infrared spectra for insulating dielectric composites with semiconductor microparticle inclusions of indium arsenide (InAs), lead sulphide (PbS), indium phosphide (InP), silicon (Si), germanium (Ge) and tellurium (Te), with direct and indirect bandgaps ranging from 0.3 to 1.4 eV. We then identify the optimal particle size of inclusions required to obtain maximal reflectance by quantifying the total reflectance from the insulating microcomposites in terms of a reflectance efficiency parameter $\eta$ for incident thermal radiation originating from black-body sources at various temperatures $T_s$. Additionally, we examine the effect of scattering from the microparticles on diffuse reflectance by comparing results from the Monte Carlo modeling with those from Fresnel equations based on the effective medium theory (EMT). The Fresnel equations take into account interference effects arising from the partial reflectance of the incident thermal radiation at the composite-ambient interfaces but do not account for scattering from the inclusions. 

\section{Theory and Methods}
For modeling thermal radiative transfer in an insulating dielectric with randomly dispersed low-bandgap semiconducting microparticles we employ a Monte Carlo method primarily developed and designed by Wang \textit{et al.} for modeling radiation transport in turbid media \cite{mcml}. To isolate the role of plasmonic resonance driven scattering in enhancing diffuse reflectance, we make the simplifying assumption that the semiconductor microparticles are embedded in an isotropic, non-scattering and non-absorbing host material with an effective dielectric constant $\epsilon_\mathrm{h}=2.25$. Additionally, the dielectric microcomposite layer is characterized by a thickness $t$, refractive index $n$, absorption coefficient $\mu_{\mathrm{abs}}$, scattering coefficient $\mu_{\mathrm{sca}}$ and a scattering anisotropy factor $g$. The composite layer is also assumed to be free-standing in a medium with a dielectric constant of $\epsilon_0 = 1$.  

\begin{figure}[h!]	
\centering
\includegraphics[width=14cm]{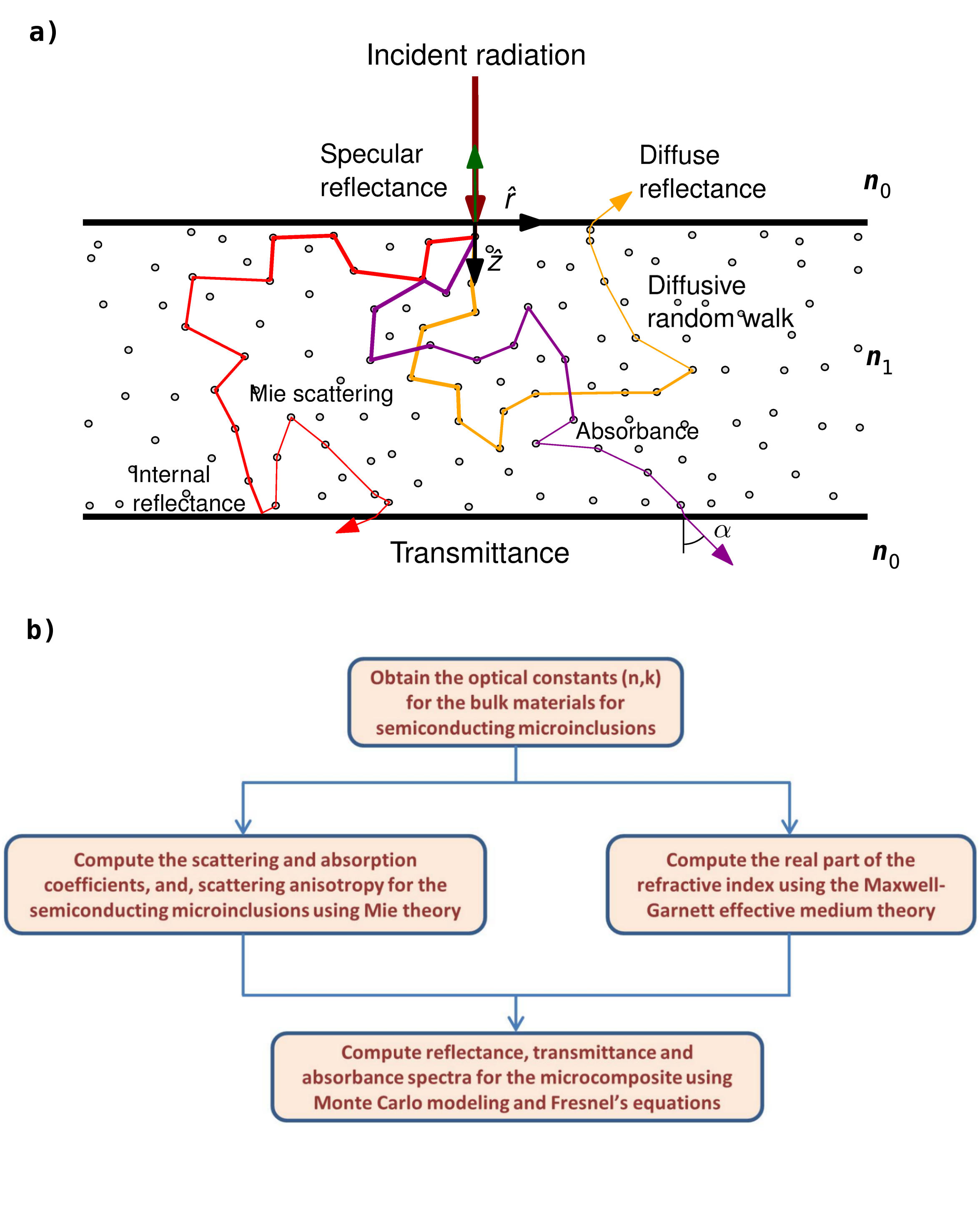}
\caption{(a) Schematic for the Monte Carlo model of propagating photons inside composites with scattering microinclusions for modeling the transport of the incident thermal radiation. An infinitesimally thin beam of incident photons is scattered within the microcomposite until either the photons are absorbed or they exit the system. The randomly distributed small open circles represent microinclusions that serve as scattering and absorption centers for the photons. The decrease in the thickness of the color trajectories represents the decrements in the photon weights as they execute random motion in the microcomposite layer. (b) Work-flow for the computation of the simulation parameters based on Mie theory and MG-EMT for use with the Monte Carlo method.}
\label{illustration} 
\end{figure}

Briefly, the Monte Carlo method models radiative thermal transport by tracking packets of energy or photons launched perpendicularly into the composite layer (See Figure \ref{illustration} for a schematic). Each photon is characterized by a weight factor that is initialized to unity before its launch. Once a photon enters the microcomposite layer, the step size $s$ for its propagation is given by

\begin{equation}
s = - \frac{\mathrm{ln} (\xi)}{\mu_{\mathrm{abs}}+\mu_{\mathrm{sca}}}.
\end{equation}

Here, $\xi$ is a random variable uniformly distributed over the interval (0,1). If during propagation the photon hits a boundary between two dissimilar media then the probability $R$ of it being reflected back is defined to be an average of the reflectances for the two orthogonal polarizations 

\begin{equation}
R=\frac{1}{2}\bigg(\frac{\mathrm{sin}^2(\phi_0-\phi_1)}{\mathrm{sin}^2(\phi_0+\phi_1)}+\frac{\mathrm{tan}^2(\phi_0-\phi_1)}{\mathrm{tan}^2(\phi_0+\phi_1)}\bigg),
\end{equation}

to account for the unpolarized nature of the incident and propagating thermal radiation. Here, $\phi_0$ and $\phi_1$ are the angles of incidence and transmittance, respectively. If the photon does not hit a boundary, its weight is decremented by the fraction of the energy absorbed in the microcomposite. A new direction is then sampled according to the Henyey-Greenstein function \cite{HG} using the scattering anisotropy $g$. The values for $g$ vary between $-1$ and $+1$ with the upper and lower limits corresponding to totally asymmetric backward and forward scattering, respectively. The photon is moved through different interaction sites in the microcomposite until it either escapes the system or its weight diminishes below $10^{-4}$ times its initial weight at the time of launching. If the photon exits the system, diffuse transmittance or reflectance, depending on the exiting direction, is incremented by the residual weight. This allows for a simultaneous computation of reflectance, transmittance and absorbance throughout a multilayer system although for our purpose we consider here only a single layer of microcomposite. 

We note here that the original Monte Carlo model developed by Wang \textit{et al.} \cite{mcml} is modified in our study to correct for the specular reflectance from the first layer that is assumed to be non-absorbing in their model. See Supporting Information (SI) for details on the modification and the validation of the modifed Monte Carlo model through a comparison with results for the optical spectra of composites obtained using the four-flux method for titanium dioxide and vanadium dioxide nanoparticle inclusions (SI Figure 1) \cite{laaksonen2014,vargas1997}. 

The effective input parameters for the microcomposite layer required for use in the Monte Carlo model are calculated using the Maxwell-Garnett effective medium theory (MG-EMT) \cite{MG} and the Mie scattering theory \cite{bohren}. This is accomplished by following the steps outlined in the flowchart shown in Figure \ref{illustration}b. Scattering and absorption coefficients per unit length $\mu_{\mathrm{sca}}$ and $\mu_{\mathrm{abs}}$ for the spherical semiconductor microparticles are calculated as

\begin{equation}
\mu_{\mathrm{sca/abs}}=\frac{3}{2}\frac{fQ_{\mathrm{sca/abs}}}{d},
\label{mu}
\end{equation}

where $f$ is the volume fraction of the particle inclusions, $d$ their diameter, and, $Q_{\mathrm{sca}}$ and $Q_{\mathrm{abs}}$ are their scattering and absorption efficiencies respectively. The efficiencies $Q_{\mathrm{sca}}$ and $Q_{\mathrm{abs}}$, in turn, are computed from the $n^{th}$ order Mie coefficients $a_n$ and $b_n$ for the electric and magnetic fields respectively using  

\begin{equation}
Q_{\mathrm{sca}} = \frac{2}{x^2} \sum^N_{n=1} (2n+1)(|a_n|^2+|b_n|^2),
\end{equation}

\begin{equation}
Q_{\mathrm{abs}} = \frac{2}{x^2} \sum^N_{n=1} (2n+1)(\mathrm{Re}(a_n+b_n) - (|a_n|^2+|b_n|^2)).
\end{equation}

Here, the size parameter $x$ is defined as the ratio of the circumference of the spherical particles to the wavelength of the incident radiation in the surrounding host medium \cite{bohren}. The order $n$ represents the various modes of the plasmonic resonance such as dipole $(n=1)$, quadrupole $(n=2)$, octupole $(n = 3)$, and so on. Detailed expressions for the Mie coefficients $a_n$ and $b_n$ can be found in Bohren and Huffman \cite{bohren}. The scattering anisotropy factor $g$ in terms of the Mie coefficients is given by

\begin{equation}
g = \frac{4}{x^2 Q_{\mathrm{sca}}} \sum^N_{n=1} \bigg[\frac{n(n+2)}{n+1}\mathrm{Re}(a_na^*_{n+1}+b_nb^*_{n+1}) + \frac{2n+1}{n(n+1)}\mathrm{Re}(a_nb^*_n)\bigg].
\end{equation}

The real part of the effective refractive index for the microcomposites is calculated from the MG-EMT formula by using the dielectric permittivities of the bulk materials comprising the host and the semiconducting microinclusions. MG-EMT approximates inhomogeneous materials as homogeneous media with effective macroscopic dielectric permittivities. The effective permittivity $\epsilon_{\mathrm{MG}}$ for a host material with spherical inclusions according to the MG formula is \cite{MG}

\begin{equation}
\epsilon_{\mathrm{MG}} = \epsilon_\mathrm{h} + 3 f \epsilon_\mathrm{h} \frac{\epsilon_\mathrm{s}-\epsilon_\mathrm{h}}{\epsilon_\mathrm{s}+2\epsilon_\mathrm{h}-f(\epsilon_\mathrm{s}-\epsilon_\mathrm{h})},
\end{equation}

where $f$ is the volume fraction of the particulate inclusions. In our simulations, the semiconducting spherical microinclusions are the sole contributors to the scattering and absorption of the incident thermal radiation in the composite layer as the host medium is non-absorbing and non-scattering. Therefore, $Q_{\mathrm{sca}}$ and $Q_{\mathrm{abs}}$ obtained from the Mie theory using an algorithm by Wiscombe \cite{wiscombe}, describe the scattering and absorption in the entire medium. 

The Maxwell-Garnett formula is based on the dipolar response of non-interacting particles to an applied electromagnetic field and its use therefore must be limited to small volume fractions $(f \leq 0.1)$ of particle inclusions. It is also well-established that classical EMTs ignore size-dependent properties of particle inclusions leaving them exclusively applicable to weakly scattering systems with particles of radii much smaller than the wavelength $\lambda$ of the incident radiation ($r < 0.1\,\lambda$) \cite{laaksonen2014}. Thus, here we use the absorption coefficients calculated using the absorption efficiencies $Q_{\mathrm{abs}} $ (Equation \ref{mu}) from the Mie theory to account for the size-dependent properties of the microinclusions in the composites in both the Monte Carlo model and the Fresnel equations. Furthermore, to understand and isolate the effect of enhanced scattering from the semicondutor microinclusions, we compare our results obtained from the Monte Carlo modeling with those computed using the Fresnel equations \cite{heavens} that account for interference effects alone.

We also define a thermal reflectance efficiency factor $\eta$ to quantify and evaluate the suitability of a given low-bandgap semiconductor material for use as microparticle inclusions in composites for thermal insulation. The efficiency factor $\eta$ describes the fraction of the incident radiation being reflected over the entire spectrum and is defined as

\begin{equation}
\eta=\frac{\int^{\lambda_1}_{\lambda_0} R(\lambda)I(T,\lambda) d\lambda}{\int^{\lambda_1}_{\lambda_0} I(T, \lambda) d\lambda}.
\label{E}
\end{equation}

where $R(\lambda)$ is the reflectance obtained from a microcomposite for a given wavelength $\lambda$. The irradiance $I(T,\lambda)$, calculated using Planck's law, corresponds to the spectral density of the electromagnetic radiation emitted by a black body source at temperature $T_\mathrm{s}$.

\section{Results}
We first examine the results from the Mie theory calculations for  the scattering $Q_{\mathrm{sca}}$ and absorption $Q_{\mathrm{abs}}$ efficiencies, and, the asymmetry factor $g$ for Ge, Si, PbS, InP, InAs and Te microparticles of various sizes $d$ followed by results from Monte Carlo modeling and Fresnel equations for the reflectance and absorbance spectra. For this study, we obtain the bulk values for the complex refractive indices of these materials from Palik \cite{palik}. For birefringent Te, the bulk refractive indices are averaged over the ordinary and the extraordinary directions. Arguably, our choice of the low-bandgap semiconductor materials for microinclusions is a \textit{priori} somewhat arbitrary. However, it is designed to understand the scattering and reflectance properties of composites with microinclusions of materials characterized by a range of direct (PbS, InAs, InP, Te) and indirect bandgaps (Si, Ge), and, elemental and compound semiconductors that are already in widespread use or are easy to synthesize in bulk using the chemical route at low cost \cite{karami, battaglia}.

\subsection{Mie scattering from semiconductor microinclusions}
\begin{figure}[H]	
\centering
\includegraphics[width=18cm,trim={7.5cm, 0cm, 7.5cm, 0cm}]{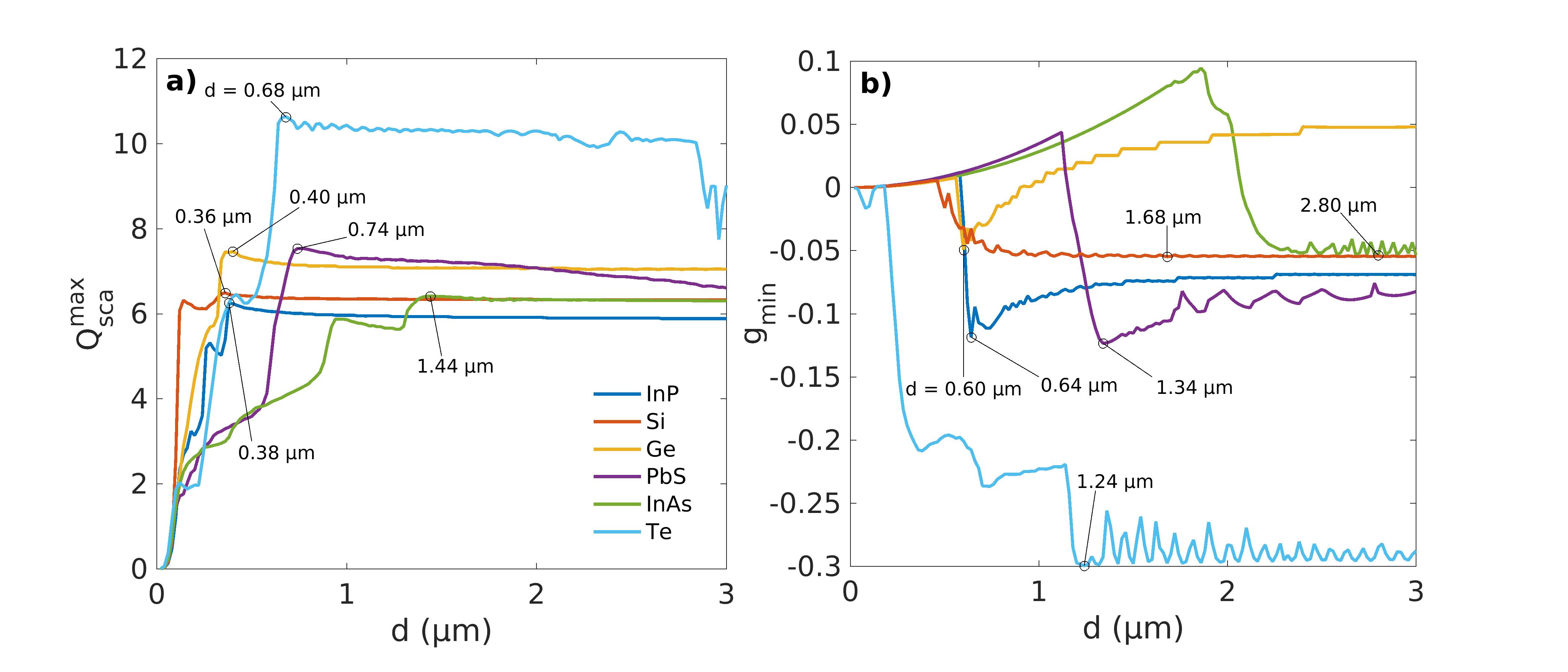}
\caption{(a) Maxima in scattering efficiency $Q_{\mathrm{sca}}$, and, (b) minima in anisotropy factor $g$ as a function of the microinclusion size $d$ for the different materials considered here. A sharp switch from forward ($+g$) to backward scattering ($-g$) with an increase particle diameter $d$ in all materials points to the presence of Fano resonances \cite{tribelsky, fan}.}
\label{Qscag} 
\end{figure}

A good microcomposite thermal insulator that minimizes radiative heat transfer should ideally maximize backscattering of the incident thermal radiation to achieve high infrared reflectance, a condition that is characterized by a high $Q_{\mathrm{sca}}$, and, a low $g$ and $Q_{\mathrm{abs}}$. For a given semiconductor material, these parameters strongly depend on the particle size $d$ and the wavelength $\lambda$ of the incident thermal radiation. Thus, we compute the Mie parameters $Q_{\mathrm{sca}}$, $Q_{\mathrm{abs}}$ and $g$ as a function of particle diameter, from $d=0.02$ to $3$ $\mathrm{\mu m}$, for wavelengths ranging from $\lambda=0.5$ to $10$ $\mathrm{\mu m}$. The maxima and minima for $Q_{\mathrm{sca}}$ and $g$ are listed in Tables \ref{tab} and \ref{tab2}, respectively, along with their corresponding wavelengths and particle sizes. Table \ref{tab} also shows the characteristic bandgap wavelengths $\lambda_{\mathrm{bg}}$ for the different semiconductor materials used as microinclusions. Figure \ref{Qscag} a-b further shows $Q^{\mathrm{max}}_{\mathrm{sca}}$ and $g_{\mathrm{min}}$ as a function of the microinclusion size $d$.

\begin{table}[H]
\centering
\caption {Values for the characteristic bandgap wavelengths $\lambda_{\mathrm{bg}}$ (indicated by vertical green arrow-marks in figures), maxima in scattering efficiency $Q^{\mathrm{max}}_{\mathrm{sca}}$ with corresponding wavelengths $\lambda_{Q_{\mathrm{sca}}^{\mathrm{max}}}$ and the microcinclusion size $d_{Q_{\mathrm{sca}}^{\mathrm{max}}}$ for the different semiconductor materials considered in this study.} 
\hfill \break
\label{tab} 
\begin{tabular}{l r r r r r r r r r}
  \hline 
  Material & $\lambda_{\mathrm{bg}}$ & $Q_{\mathrm{sca}}^{\mathrm{max}}$ & $d_{Q_{\mathrm{sca}}^{\mathrm{max}}}$ & $\lambda_{Q_{\mathrm{sca}}^{\mathrm{max}}}$ \\
  & $\mathrm{\mu m}$ & & $\mathrm{\mu m}$ & $\mathrm{\mu m}$ \\
  \hline 
  InP & 0.92 & 6.3 & 0.38 & 0.95 \\
  Si & 1.11 & 6.5 & 0.36 & 0.97 \\
  Ge & 1.85 & 7.5 & 0.40 & 1.8 \\ 
  PbS & 3.35 & 7.5 & 0.74 & 3.4 \\
  InAs & 3.44 & 6.4 & 1.44 & 3.8 \\
  Te & 3.75 & 10.6 & 0.68 & 4.0 \\
  \hline 
\end{tabular}
\centering 
\end{table}

\begin{figure}[H]	
\centering
\includegraphics[width=12.5cm, height=19.8cm, trim={8cm, 0cm, 8cm, 0cm}]{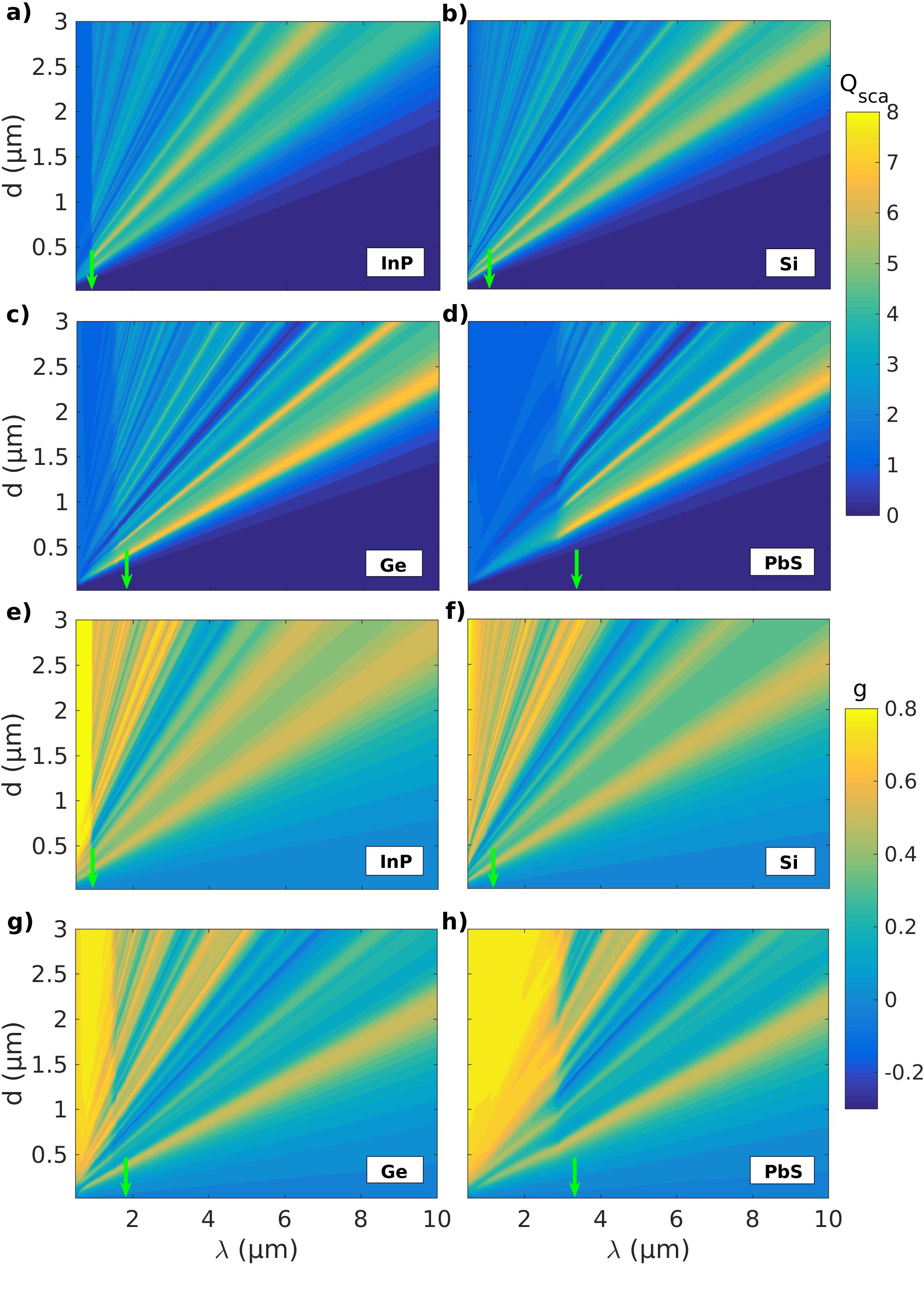}
\caption{(a-d) Scattering efficiency $Q_{\mathrm{sca}}$, and, (e-h) anisotropy factor $g$ as a function of the wavelength $\lambda$ of the incident thermal radiation and the diameter $d$ of spherical InP, Si, Ge and PbS microinclusions, respectively. The bandgap wavelengths $\lambda_{\mathrm{bg}}$ (indicated by vertical green arrow-marks) for the semiconductor materials mark a transition from low to high $Q_{\mathrm{sca}}$ and strongly forward ($+g$) to mixed scattering regimes for the microinclusions with an increase in $\lambda$.} 
\label{Qscag2D} 
\end{figure}

Absorption of the incident thermal radiation at wavelengths close to the absorption band-edge $(\lambda \approx \lambda_{\mathrm{bg}}$, Table \ref{tab}, $\lambda_{\mathrm{bg}}$ indicated by vertical green arrow-marks on the x-axis in figures.) gives rise to a significant increase in the number of charge carriers in the conduction (electrons) or the valence band (holes) leading to the excitation of plasmonic resonances in the semiconducting microinclusions. These resonances result in the formation of oscillating multipoles that radiate to generate large values of $Q_{\mathrm{sca}}$ characterized by broad maxima as shown in Figure \ref{Qscag2D}a-d and SI Figure 2a-b. Figure \ref{Qscag2D}a-d also shows that the maxima in $Q_{\mathrm{sca}}$ occur when the wavelength of the incident radiation is comparable to the size $d$ of the microparticles. For particle sizes $d \leq 0.1$ $\mathrm{\mu m}$, $Q_{\mathrm{sca}}$ remains well below 2.6 for all microinclusion materials and does not attain large values for $\lambda \leq \lambda_{\mathrm{bg}}$ as seen in Figure \ref{Qscag2D}. This behavior is particularly apparent for composites with PbS (Figure \ref{Qscag2D}d), InAs and Te (SI Figure 2a-b) microinclusions that have small bandgaps. Figure \ref{Qscag}a shows that (i) $Q_{\mathrm{sca}}$ attains a maxima at smaller particle sizes for microinclusions of semiconductors with larger bandgaps or smaller $\lambda_{\mathrm{bg}}$ (vertical green arrow-marks), and, (ii) after the maxima is attained, $Q_{\mathrm{sca}}$ remains more or less constant with any further increase in particle size. Also, particles with sizes comparable to the wavelength of the incident thermal radiation exhibit strong forward scattering ($g>0$) for $\lambda < \lambda_{\mathrm{bg}}$ (Figure \ref{Qscag2D} e-h and SI Figure 2c-d). However, in the limit of Rayleigh scattering the small nanoscale particles exhibit isotropic scattering characterized by $g$ values close to zero. 

\begin{figure}[H]	
\centering
\includegraphics[width=14cm,trim={6.5cm, 0cm, 6.5cm, 0cm}]{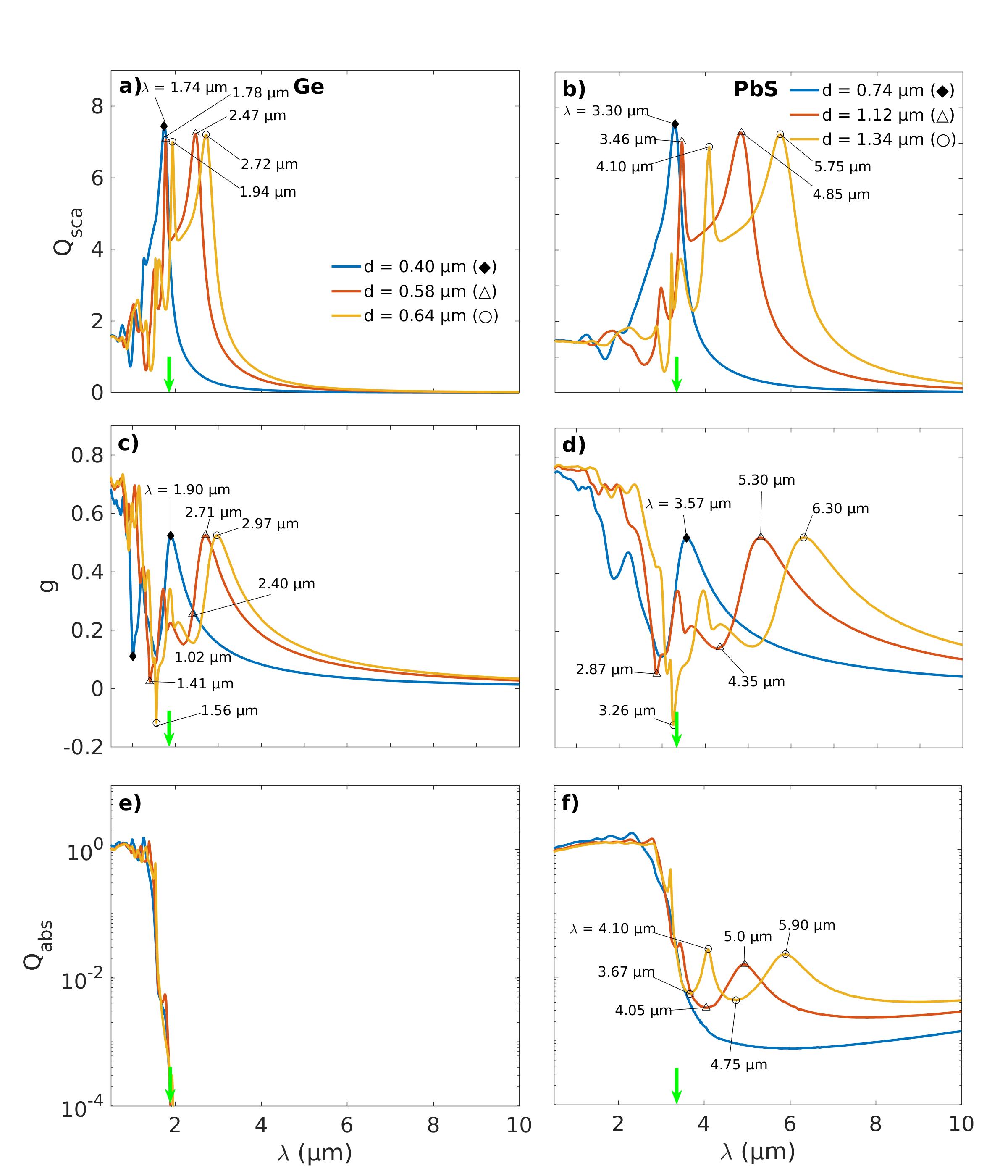}
\caption{Mie parameters: (a, b) Scattering efficiencies $Q_{\mathrm{sca}}$, (c, d) scattering anisotropy $g$, and,  (e, f) absorption efficiencies $Q_{\mathrm{abs}}$ for various sizes of Ge (left) and PbS (right) microinclusions. The vertical green arrows indicate the bandgap wavelength $\lambda_{\mathrm{bg}}$ for the semiconductor materials. A general broadening of the spectral features in $Q_{\mathrm{sca}}$, $Q_{\mathrm{abs}}$ and $g$ is observed with an increase in the microinclusion size $d$.}
\label{Mie} 
\end{figure}

Furthermore, it is observed that the local maxima in $Q_{\mathrm{sca}}$ and plot features in $g$ redshift and broaden as the particle size is increased for all semiconducting microinclusion materials considered here (Figures $\ref{Qscag2D}$, \ref{Mie}a-d, and, SI Figures 2 and 3). This occurs for increased particle sizes because of a weakening of the restoring force that drives the plasmonic resonances. The restoring force weakens due to an increased distance between the oscillating charges on the opposite sides of a particle leading to a consequent weakening of the interaction between them and hence lower associated energies or a redshift. The effect can be seen more readily when the spectral behavior of $Q_{\mathrm{sca}}$ and $g$ is plotted for Ge and PbS in Figure \ref{Mie}(a-d) for different particle sizes corresponding to $Q_{\mathrm{sca}}^{\mathrm{max}}$ and $g_{min}$ shown in Figure \ref{Qscag} and Tables \ref{tab} and \ref{tab2}. For example Figure \ref{Mie}a shows that the peaks in $Q_{\mathrm{sca}}$ for Ge at $\lambda={1.78}$ and $2.47$ $\mathrm{\mu m}$ redshift to $\lambda={1.93}$ and $2.72$ $\mathrm{\mu m}$ when the particle size increases from $d=0.58$ to $0.64$ $\mathrm{\mu m}$ ($\triangle$, $\circ$). Similar shifts are observed in $g$ for Ge (Figure \ref{Mie}c), and, $Q_{\mathrm{sca}}$ and $g$ for PbS in Figure \ref{Mie}b and d respectively. 

\begin{figure}[H]	
\centering
\includegraphics[width=14cm,height=19.7cm,trim={7.5cm, 0cm, 7cm, 0cm}]{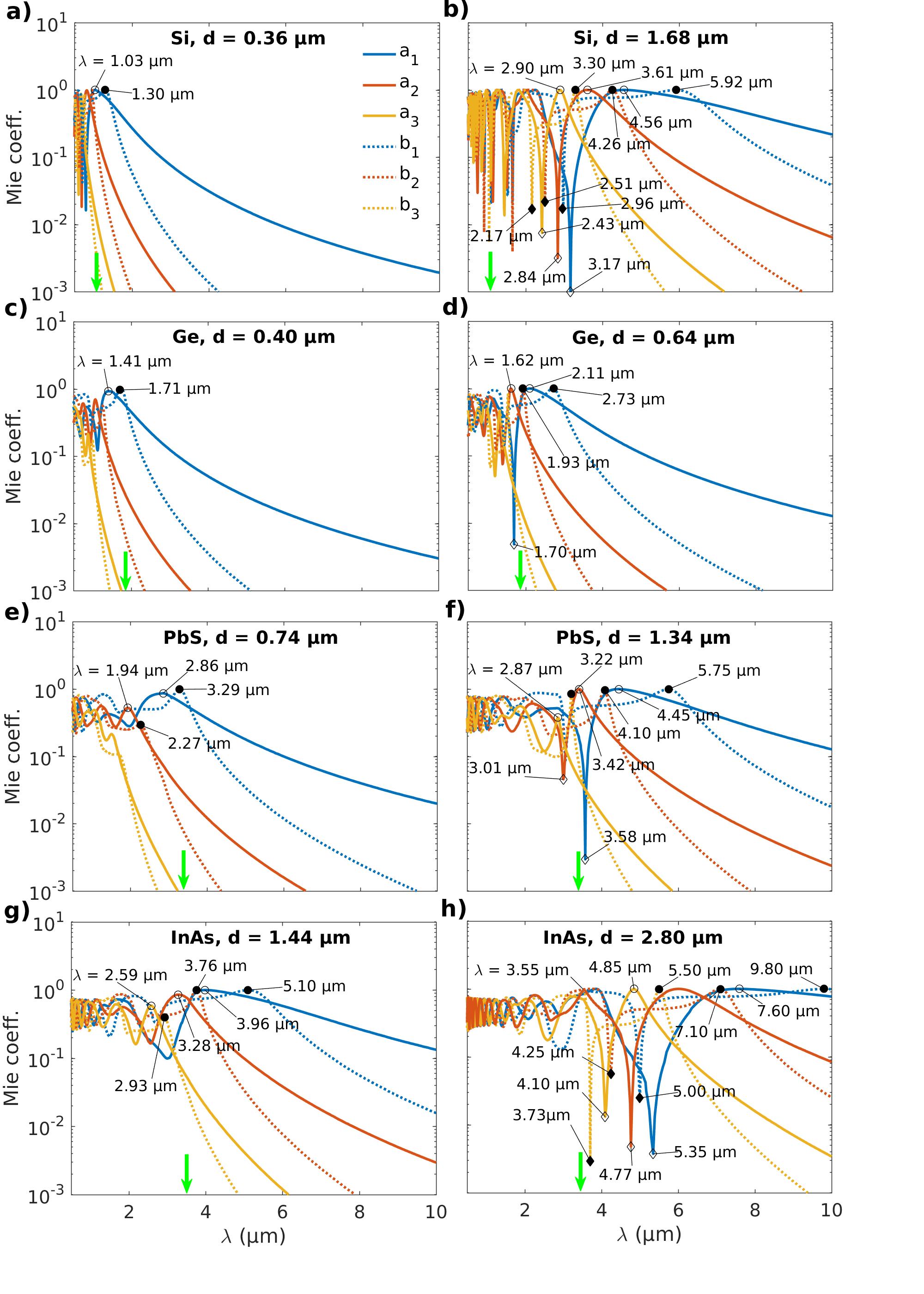}
\caption{Mie coefficients $a_{n}$ and $b_{n}$ as a function of wavelength $\lambda$ for spherical microinclusions of (a, b) Si, (c,d) Ge, (e, f) PbS and (g, h) InAs for particle diameters $d$ corresponding to $Q^{\mathrm{max}}_{\mathrm{sca}}$ and $g_{\mathrm{min}}$ as shown in Figure \ref{Qscag}. The vertical green arrows indicate the bandgap wavelengths $\lambda_{\mathrm{bg}}$. An increase in the microinclusion size $d$ is accompanied by a strengthening of the magnetic modes $b_n$. Open and closed symbols denote features in $a_n$ and $b_n$ respectively.}
\label{MieABn} 
\end{figure}

Figure \ref{MieABn} and SI Figure 4 present the Mie coefficients $a_n$ and $b_n$ for the particles of different semiconductors with sizes $d$ corresponding to $Q^{\mathrm{max}}_{\mathrm{sca}}$ and $g_{\mathrm{min}}$. Compared to the dipole modes, it is observed that the Mie coefficients for the quadrupole and octupole modes decay much faster with increasing wavelength of the incident thermal radiation. As a result, one need only consider the first three modes of the Mie coefficients $a_n$ ($\circ$, $\diamond$) and $b_n$ ($\bullet$, $\scriptstyle\bigLozenge$) \textit{i.e.} dipole, quadrupole and octupole. Consistent with the features in plots for $Q_{\mathrm{sca}}$ and $g$ (Figures \ref{Qscag2D}, \ref{Mie}, and, SI Figures 2 and 3), the plasmonic resonances ($\circ$, $\bullet$) are seen to broaden and red-shift with an increase in the semiconductor particle size $d$ (Figure \ref{MieABn} and SI Figure 4). Sharp dips ($\diamond$, $\scriptstyle\bigLozenge$) in the values of the Mie coefficients indicate minima in the extinction efficiency ($Q_{\mathrm{ext}}=Q_{\mathrm{sca}}+Q_{\mathrm{abs}}$) and consequently an increase in transmittance. Results also indicate that the magnetic Mie modes are weaker and decay much faster than the electric modes for all the particle sizes and semiconductor materials considered here (Figure \ref{MieABn} and SI Figure 4). However, consistent with theoretical predictions, a strengthening of the magnetic modes $b_n$ is observed with an increase in the particle size \cite{fanSingh}. This strengthening of the magnetic modes is much greater for the Si, PbS, InAs and Te microparticles (Figure 6a-b, e-h and SI Figure 4c-d) compared to that for Ge or InP inclusions (Figure \ref{MieABn}c-d and SI Figure 4a-b). Also, the sharp quadrupole and octupole resonances occurring against a background of broad dipole modes for the larger particles give rise to Fano resonances as evidenced by an abrupt switch in the scattering anisotropy $g$ from forward ($g>0$) to backward scattering ($g<0$) with an increase in particle size (Figure \ref{Qscag}b) \cite{tribelsky, fan}. 

Further, sharp resonances in $Q_{\mathrm{sca}}$ for the semiconducting microinclusions can largely be attributed to the points in the spectra where the Mie coefficients $a_n$ ($\circ$) and $b_n$ ($\bullet$) for the electric and magnetic fields, respectively, tend to unity (or maxima), a condition required for the occurrence of scattering resonances \cite{tribelsky}. Again, considering Ge and PbS as illustrative examples, it can be seen that there occur Fano resonances in $Q_{\mathrm{sca}}$ at $\lambda=1.93$ and $2.72$ $\mathrm{\mu m}$ for Ge particles of size $d=0.64$ $\mathrm{\mu m}$ ($\circ$) (Figure \ref{Mie}a), and, at $\lambda={4.10}$ and $5.75$ $\mathrm{\mu m}$ for PbS particles of size $d=1.34$ $\mathrm{\mu m}$ ($\circ$) (Figure \ref{Mie}b). These strong resonances in $Q_{\mathrm{sca}}$ (Figure \ref{Mie}a-b) can be attributed to the sharp maxima occuring at the same or close wavelengths in the Mie coefficients $b_1$ and $b_2$ corresponding to the magnetic field against a background of the broad contribution to scattering from the electric dipole mode $a_1$ (Figure \ref{MieABn}c-f). For Si microinclusions of size $d=1.68$ $\mathrm{\mu m}$ multiple sharp maxima are seen for the dipole, quadrupole and octupole modes for both electric and magnetic Mie coefficients resulting in a large number of Fano resonances in $Q_{\mathrm{sca}}$ (Figures \ref{MieABn}a-b and SI Figure 3b respectively). Similar correspondence between the maxima in $a_n$, $b_n$ and the peaks in $Q_{\mathrm{sca}}$ occurs for InP (SI Figures 4a-b and 3a), InAs (Figure \ref{MieABn}g-h and SI Figure 3c) and Te (SI Figures 4c-d and 3d ) microinclusions as well. However, more generally, specific features in $Q_{\mathrm{sca}}$ and $g$ arise from interference effects among the Mie coefficients of different orders.

At the absorption band edge marked by $\lambda_{\mathrm{bg}}$ (Table \ref{tab}), a steep increase in $Q_{\mathrm{abs}}$ is observed with decreasing $\lambda$ for particles of all materials (Figure \ref{Mie}e-f and SI Figures 5 and 6). The resonances in Mie coefficients $a_n$ and $b_n$ extend beyond $\lambda_{\mathrm{bg}}$ for all materials but $Q_{\mathrm{abs}}$ essentially goes to zero outside the main absorption band, as is to be expected, only for the Ge (Figure \ref{Mie}e and SI Figure 5c), InP and Si microinclusions (SI Figures 5a-b and 6a-b).  However, broad peaks in $Q_{\mathrm{abs}}$ that exist far away from the main absorption band at longer wavelengths and are about $10-20$ times weaker are seen for PbS (Figure \ref{Mie}f), InAs and Te microinclusions (SI Figures 6c-d). These distinctive long-wavelength absorption bands broaden and move farther away from the main absorption band with an increase in the microinclusion size $d$. This is seen in $Q_{\mathrm{abs}}$ for PbS particles presented in Figure \ref{Mie}f where these bands with peaks at $\lambda=4.10$ and $5.90$ $\mathrm{\mu m}$ become distinctive for particles of diameter $d=1.34$ $\mathrm{\mu m}$ ($\circ$). Correspondingly, peaks are also observed in $Q_{\mathrm{sca}}$ along with associated features in $g$ and the Mie coefficients $a_n$ and $b_n$ at close wavelengths, as described earlier (Figures \ref{Mie}b,d and \ref{MieABn}e-f, respectively). This, therefore, points to the generation of a sufficiently large number of free charge carriers at $\lambda > \lambda_{\mathrm{bg}}$ to enable the generation of plasmonic resonances. Also, it appears that the origin of the weak absorption peaks in $Q_{\mathrm{abs}}$ for PbS (Figure \ref{Mie}f), InAs (SI Figure 6c) and Te (SI Figure 6d) microcomposites is likely due to a cluster of defect states within the bandgap with intermediate energies corresponding to the incident thermal radiation. These weak absorption bands at longer wavelengths  ($\lambda > \lambda_{\mathrm{bg}}$) serve to extend maxima in $Q_{\mathrm{sca}}$ much beyond the absorption band-edge (Figure \ref{Mie}b and SI Figure 3c-d). However, in the absence of any significant absorption away from the main absorption band (SI Figure 6b), the origin of the several peaks observed in the spectra of $Q_{\mathrm{sca}}$ for Si microinclusions of size $d =1.68$ $\mathrm{\mu m}$ ($\circ$) is an exception (SI Figure 3b). This may, however, be a result of the complex nature of the band-structure for Si and its indirect bandgap, a discussion of which is beyond the scope of the current article. 

\subsection{Spectral reflectance of microcomposites}
This section presents results on the spectral characteristics of composites with low-bandgap semiconductor microinclusions computed using Monte Carlo modeling and Fresnel equations. For Monte Carlo modeling, we employ the spherical microinclusions of optimal size $d$  determined using Mie theory for obtaining maximum $Q_{\mathrm{sca}}$ and minimum $g$ for the various semiconductor materials (Table \ref{tab}). Furthermore, for all our computations here, we consider a microcomposite with a thickness $t=200$ $\mathrm{\mu m}$ and a semiconductor microinclusion volume fraction of $f = 0.01$ unless specified otherwise. Considering a cylindrical symmetry for the propagation of the infinitesimally thin beam of incident thermal radiation in the Monte Carlo model, a grid resolution of $dz = 2$ $\mathrm{\mu m}$ and $dr = 1$ $\mathrm{\mu m}$ is used for the radial $\hat{r}$ and axial $\hat{z}$ directions respectively (see Figure \ref{illustration}). The total number of grid elements in the $\hat{r}$-direction is set to $N_r=100$ while the number of grid elements $N_z$ in the $\hat{z}$-direction is determined by the thickness of the microcomposite layer. Adequate care is also taken to ensure that the diffuse reflectance and transmittance go to zero as a function of the radius $r$ while their angular dependence on the photon-exiting direction $\hat{\alpha}$ is ignored. To compute the infrared spectra for the incident thermal radiation, $10^7$ photons are launched for each wavelength $\lambda$ considered. 

\begin{figure}[H]	
\centering
\includegraphics[width=15cm,trim={8cm, 0cm, 8cm, 0cm}]{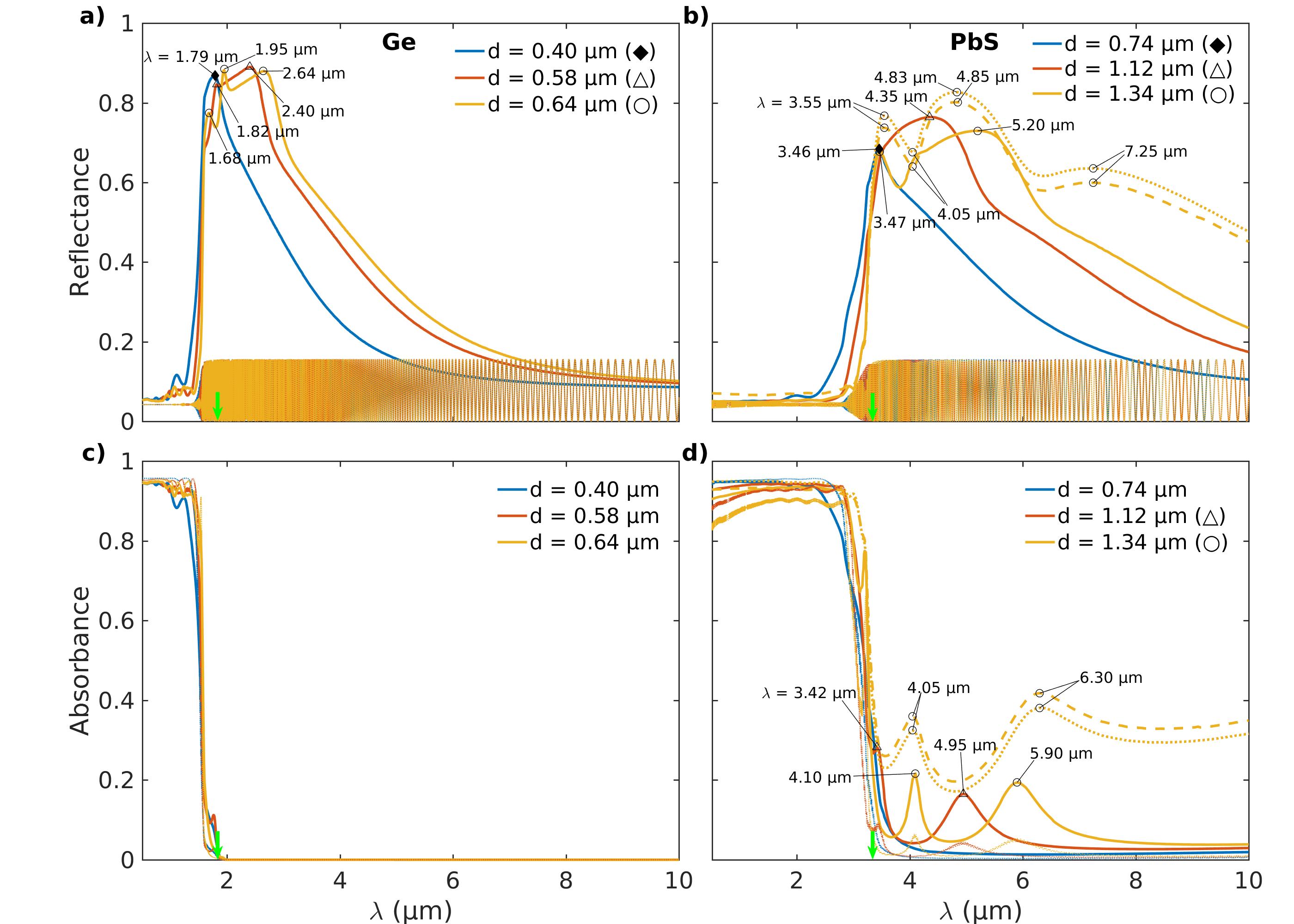}
\caption{The spectral reflectance and absorbance of microcomposites with (a, c) Ge and (b, d) PbS spherical inclusions of diameter $d$ and volume fraction $f=0.01$, respectively. The solid lines and the thin dotted lines of the same color represent spectral results obtained from the Monte Carlo modeling and Fresnel equations, respectively. In (b, d), the additional dashed and thick dotted curves in yellow color correspond to results computed using Monte Carlo modeling with a microinclusion volume fraction of $f = 0.1$ and a microcomposite of thickness 2 mm, respectivley. The green arrows indicate the bandgap wavelengths $\lambda_{\mathrm{bg}}$. In (b, d), a broadbanding of the reflectance spectra can be attributed to the plasmonic resonances arising from the collective oscillations of the free charge carriers generated due to weak absorption bands away from the absorption band edge for $\lambda > \lambda_{\mathrm{bg}}$.}
\label{GePbSRA} 
\end{figure}

Figures \ref{GePbSRA} - \ref{SiTeRAnm} show the reflection and the absorption spectra for infrared radiation ranging from $\lambda=0.5$ to $10$ $\mathrm{\mu m}$ for composite layers with Ge and PbS, and, Si and Te microinclusions, respectively.  A comparison of the results from Monte Carlo modeling and Fresnel equations for radiation transport clearly shows that the presence of the low-bandgap semiconducting microinclusions significantly increases both the reflectance and the absorbance of the microcomposite layers (Figure \ref{GePbSRA} and SI Figure 7). This is because, unlike Fresnel equations, the Monte Carlo model takes into account the plasmonic resonance induced enhanced scattering from the microparticles. This results in a decreased mean free path $(\propto [\mu_{\mathrm{abs}} +\mu_{\mathrm{sca}}]^{-1})$ and diffusive transport of the incident radiation in the microcomposite layer thereby giving rise to greater absorbance and reflectance. For a host medium refractive index of $n_m=1.5$, among the semiconductor materials considered, the highest reflectance $R=0.91$ is obtained for Te microcomposites at $\lambda=4.0$ $\mathrm{\mu m}$ for microinclusions of size $d=0.68$ $\mathrm{\mu m}$ ($\scriptstyle\bigLozenge$) (Figure \ref{SiTeRAnm}b). A similar value of $R=0.90$ is also obtained for the Si microcomposites at $\lambda=1.27$ $\mathrm{\mu m}$ for inclusions of diameter $d=0.36$ $\mathrm{\mu m}$ ($\scriptstyle\bigLozenge$) (Figure \ref{SiTeRAnm}a). Furthermore, for microcomposites with Ge inclusions of diameter $d=0.64$ $\mathrm{\mu m}$ ($\circ$) (Figure \ref{GePbSRA}a), two high peaks ($R\approx0.88$) in the reflectance ocurring at $\lambda=1.95$ and $2.64$ $\mathrm{\mu m}$ can be directly attributed to the peaks in $Q_{\mathrm{sca}}$ at $\lambda=1.94$ and $2.72$ $\mathrm{\mu m}$ (Figure \ref{Mie}a). On the other hand, the reflectance calculated using Fresnel equations for all microcomposites remains well below $R=0.2$ (Figures \ref{GePbSRA}a-b and SI Figure 7a-b). This difference between the results from Monte Carlo modeling and Fresnel equations emphasizes the hugely disproportionate impact a small volume fraction of microparticle inclusions makes on the infrared spectra of the micromposite layer. Additionally, they also underline the importance of considering scattering from particles that are comparable in size to the wavelength $\lambda$ of the incident radiation.

The peaks in reflectance are seen to redshift and broaden by various amounts for the different microcomposites with an increase in the size $d$ of the particle inclusions (see Figures \ref{GePbSRA}a-b, \ref{SiTeRAnm}a-b and SI Figure 7a-b). The effect is observed to be especially pronounced for composites with PbS, Si, Te and InAs microinclusions (Figures \ref{GePbSRA}b, \ref{SiTeRAnm}a-b, SI Figure 7b, respectively). This broadbanding of the reflectance spectra is a direct consequence of the red-shifting and broadening of the sharp Fano resonances for larger microinclusions in the spectra for $Q_{\mathrm{sca}}$ (Figure \ref{Mie}b and SI Figure 3b-d). In PbS, Te and InAs microcomposites (Figures \ref{GePbSRA}b, \ref{SiTeRAnm}b and SI Figure 7b, respectively), the broadbanding of the reflectance for the larger microinclusions appears to be driven, in part, by the enhanced scattering from plasmonic resonances generated due to the presence of weak absorption peaks far outside the main absorption band (Figures \ref{GePbSRA}d, \ref{SiTeRAnm}d and SI Figure 7d, respectively). It is notable in this regard that Felts \textit{et al.} \cite{felts} have experimentally observed LSPRs in silicon-doped InAs microparticles of size $1.0$ $\mathrm{\mu m}$ with characteristic absorbance at wavelengths $\lambda=5.75$ and $7.70$ $\mathrm{\mu m}$. The wavelengths at which these LSPR-associated absorbance maxima occur are similar to the wavelengths we observe for the absorbance maxima in composites with InAs microinclusions at $\lambda=5.55$ and $7.15$ $\mathrm{\mu m}$ ($\circ$), and, $\lambda=5.35$ $\mathrm{\mu m}$ ($\diamond$) for particles of size $d=2.80$ and $1.44$ $\mathrm{\mu m}$, respectively ($n_m=1.5$, SI Figure 7d). Additionally, it is also observed that microcomposites with larger inclusions exhibit lower maxima in reflectance (Figures \ref{GePbSRA}a-b, \ref{SiTeRAnm}a-b and SI Figure 7a-b), although the maxima in $Q_{\mathrm{sca}}(\lambda, d)$ remain approximately constant with any further increase in $d$ after they reach a peak value (Figure \ref{Qscag}a). This happens because, for a given volume fraction $f$, the scattering coefficient $\mu_{\mathrm{sca}}$ in Equation (\ref{mu}) is directly proportional to $Q_{\mathrm{sca}}$ but scales inversely with $d$. 

In all the microcomposites studied here, plasmonic resonance driven peaks in reflectance spectra (Figures \ref{GePbSRA}a-b, \ref{SiTeRAnm}a-b, SI Figure 7a-b) appear right before the absorption band edge due to low characteristic values of $Q_{\mathrm{abs}}$ for wavelengths $\lambda \gtrsim \lambda_{\mathrm{bg}}$ (Figure \ref{Mie}e-f, SI Figures 5 and 6). This is regardless of whether there exists a maxima in $Q_{\mathrm{sca}}(\lambda, d)$ or not in that wavelength range for a given microinclusion size. This is illustrated by microcomposites with PbS particles of diameter $d=1.34$ $\mathrm{\mu m}$ ($\circ$) that present a peak in reflectance with $R=0.68$ at $\lambda=3.47$ $\mathrm{\mu m}$ in Figure \ref{GePbSRA}b despite the moderate $Q_{\mathrm{sca}}=3.63$ and a value of $g=7.63 \cdot 10^{-2}$ pointing to isotropic scattering (Figure \ref{Mie}b,d). On the other hand, comparable scattering parameters $Q_{\mathrm{sca}} = 3.90$ and $g=3.60 \cdot 10^{-2}$ at $\lambda=3.22$ $\mathrm{\mu m}$ (Figure \ref{Mie}b,d) suggest higher reflectance although the actual observed reflectance $R=0.24$ is quite low compared to $R=0.68$ (Figure \ref{GePbSRA}b). Still, a significant change in reflectance occurs due to the absorption efficiency decreasing from $Q_{\mathrm{abs}} = 0.48$ at $\lambda=3.22$ $\mathrm{\mu m}$ to a low value of $Q_{\mathrm{abs}} = 8.70 \cdot 10^{-2}$ at $\lambda=3.47$ $\mathrm{\mu m}$ (Figure \ref{Mie}e).

Figure \ref{GePbSRA}a,b shows that the reflectance values $R=0.71, 0.52$ associated with Ge and PbS microinclusions of size $d=0.64$, $1.34$ $\mathrm{\mu m}$ and corresponding to the minima in scattering anisotropy $g_{\mathrm{min}}=(-1.19, 1.23)\cdot 10^{-1}$ at $\lambda=1.56$, $3.26$ $\mathrm{\mu m}$, respectively (Figure \ref{Qscag}b, Table \ref{tab2}), are not the highest values of reflectance obtained for both Ge and PbS. In the case of Ge and PbS microinclusions, this is in part explained by the fact that the wavelengths $\lambda_{g_{\mathrm{min}}}$ (Table \ref{tab2}) corresponding to $g_{\mathrm{min}}$ (Figures \ref{Mie}c-d) are located within the main absorption band (Table \ref{tab}) wherein $Q_{\mathrm{sca}}$ is low  (Figure \ref{Mie}a-b) and $Q_{\mathrm{abs}}$ is high (Figure \ref{Mie}e-f). Furthermore, both Ge and PbS microinclusions of sizes $d = 0.58, 1.12$ $\mathrm{\mu m}$ ($\triangle$) are found to be forward-scattering for the reflectance maxima at $\lambda=2.40, 4.35$ $\mathrm{\mu m}$ (Figure \ref{GePbSRA}a-b) with scattering anisotropy $g=0.25, 0.14$ (Figure \ref{Mie}c-d), respectively, thereby implying that a low value of the scattering anisotropy $g$ is not essential to obtain high reflectance. Composites with InAs microinclusions of size $d_{g_{\mathrm{min}}}=2.80$ $\mathrm{\mu m}$ show a reflectance $R_{g_{\mathrm{max}}}=0.57$ that is higher than the reflectance $R_{g_{\mathrm{min}}}=0.50$ (Table \ref{tab2}). On the other hand, as per expectations, composites with InP, Si and Te microinclusions of size $d_{g_{\mathrm{min}}}$ exhibit a higher reflectance $R_{g_{\mathrm{min}}}$ than $R_{g_{\mathrm{max}}}$ (Table \ref{tab2}). Thus, there appears to be scant correlation between a low negative value for the scattering anisotropy $g$ and a high value of reflectance $R$ due to the conflicting evidence presented by the results for the microinclusion materials considered here. This is likely because once a photon is launched into a highly scattering microcomposite layer, early on during its motion, the direction of propagation of the photon gets quickly randomized. As a consequence, a low negative value of the scattering anisotropy $g$ is rendered rather ineffective compared to the stronger influence of the scattering ($Q_{\mathrm{sca}}$) and absorption ($Q_{\mathrm{abs}}$) efficiencies.

\begin{table}[H]
\centering
\caption {Maxima and minima in the scattering anisotropy $g$ for composites with microinclusions of size $d_{g_{\mathrm{min}}}$ along with corresponding reflectances $R_{g_{\mathrm{max}}}$ and $R_{g_{\mathrm{min}}}$ at wavelengths $\lambda_{g_{\mathrm{max}}}$ and $\lambda_{g_{\mathrm{min}}}$ respectively.} 
\hfill \break
\label{tab2} 
\begin{tabular}{l r r r r r r r r}
  \hline 
  Material  & $g_{\mathrm{min}}$ & $d_{g_{\mathrm{min}}}$ & $\lambda_{g_{\mathrm{min}}}$ & $R_{g_{\mathrm{min}}}$ & $g_{\mathrm{max}}$ & $\lambda_{g_{\mathrm{max}}}$ & $R_{g_{\mathrm{max}}}$ \\
  & & $\mathrm{\mu m}$ & $\mathrm{\mu m}$ & & & $\mathrm{\mu m}$ & \\
  \hline 
  InP  & -4.94 $\cdot 10^{-2}$ & 0.60 & 0.960 & 0.76 & 0.547 & 2.16 & 0.71 \\
  Si & -5.48 $\cdot 10^{-2}$ & 1.68 & 2.87 & 0.63 & 0.711 & 1.21 & 0.49\\
  Ge & -1.19 $\cdot 10^{-1}$ & 0.64 & 1.56 & 0.71 & 0.524 & 2.98 & 0.72 \\ 
  PbS & -1.23 $\cdot 10^{-1}$ & 1.34 & 3.26 & 0.52 & 0.523 & 6.30 & 0.54 \\
  InAs & -5.35 $\cdot 10^{-2}$ & 2.80 & 4.90 & 0.50 & 0.451 & 6.89 & 0.57 \\
  Te & -2.99 $\cdot 10^{-1}$ & 1.24 & 3.92 & 0.65 & 0.513 & 7.85 & 0.51 \\
  \hline 
\end{tabular}
\centering 
\end{table}

Figure \ref{GePbSRA}b,d shows the reflectance and absorbance spectra for the microcomposites with PbS microinclusions of diameter $d=1.34$ $\mathrm{\mu m}$ ($\circ$) for two different volume fractions $f = 0.01$ ($t=200$ $\mathrm{\mu m}$ and $\SI{2}{\milli\meter}$) and $0.1$. It is observed that the increase in volume fraction from $f = 0.01$ to $0.1$ shifts the peak in reflectance at $\lambda=5.20$ $\mathrm{\mu m}$ to $\lambda=4.85$ $\mathrm{\mu m}$ and results in a new reflectance peak at $\lambda=7.25$ $\mathrm{\mu m}$. The peak at $\lambda=7.25$ $\mathrm{\mu m}$ also appears in the reflectance for the microcomposite with a PbS particle volume fraction $f = 0.01$ and thickness $t=\SI{2}{\milli\meter}$. More generally, this implies that a larger number of particles is required to produce enough scattering to reflect the longer wavelength infrared radiation because a microcomposite of thickness $t=200$ $\mathrm{\mu m}$ and volume fraction $f = 0.01$ has only ${1/10}^{th}$ the number of particles compared to the other two microcomposites with increased thickness ($t=\SI{2}{\milli\meter}$) and volume fraction ($f=0.1$), respectively.

An increase in the volume fraction $f$ of the low-bandgap semiconducting microinclusions increases scattering and hence has the general effect of increasing the reflectance $R$ of the microcomposite. However, beyond a point any further increase in $f$ to increase $R$ is counteracted by an increase in the absorbance that would be detrimental to the performance of an insulating microcomposite. This is evident from Figure \ref{GePbSRA}b wherein the reflectance at $\lambda=4.05$ $\mathrm{\mu m}$ for a PbS microcomposite decreases from a value of $R=0.68$ for $f = 0.01$ ($t=200$ $\mathrm{\mu m}$) to $R=0.64$ for $f = 0.1$ ($t=200$ $\mathrm{\mu m}$). 

\subsection{Nature of plasmonic resonances}
\begin{figure}[H]	
\centering
\includegraphics[width=13.5cm,trim={7cm, 0cm, 6cm, 0cm}]{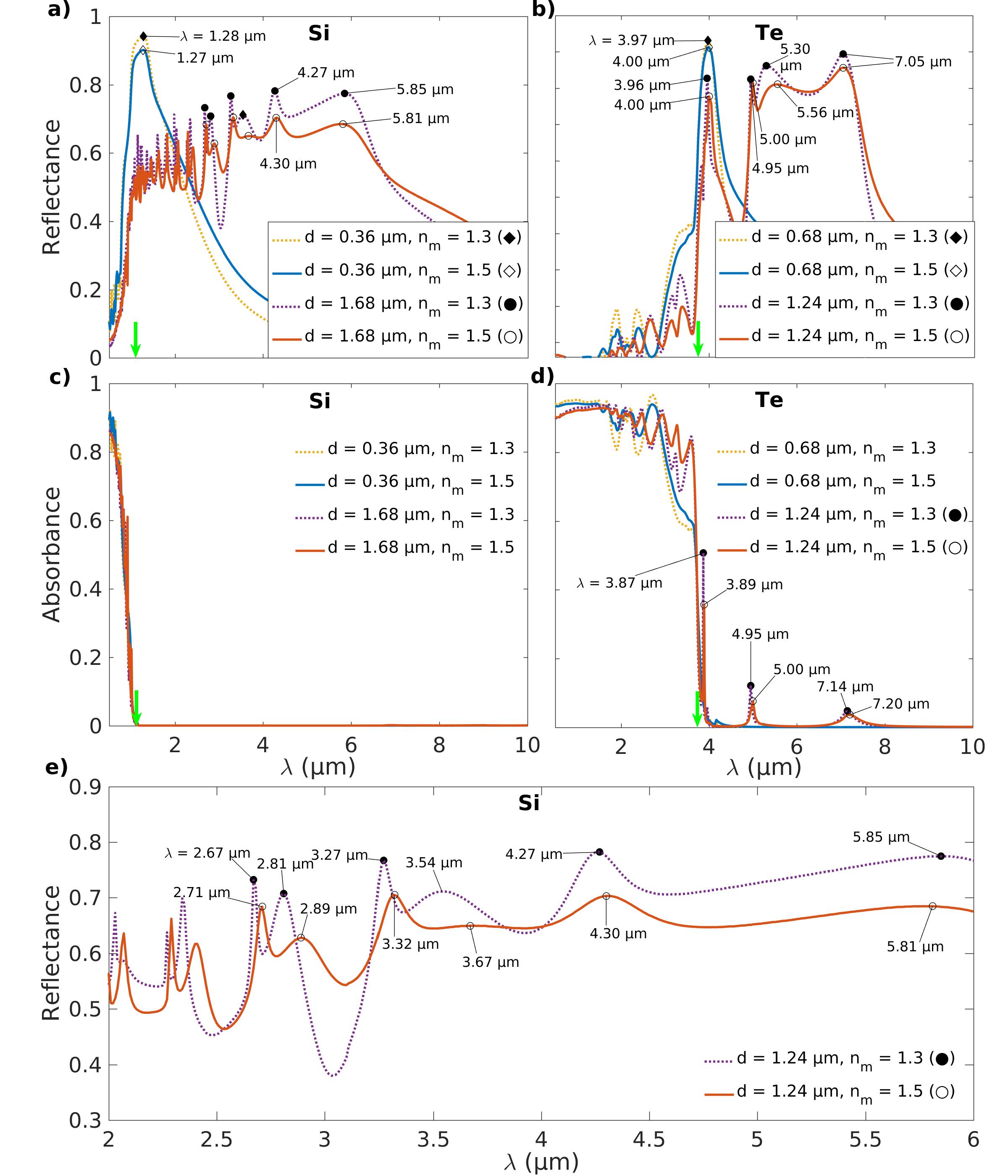}
\caption{(a-b) Spectral reflectance and (c-d) absorbance for microcomposites with a volume fraction $f=0.01$ of (a, c) Si and (b, d) Te particles of different sizes $d$ embedded in a dielectric medium of refractive index $n_m = 1.5$ and $1.3$. (e) An expanded view of the reflectance peaks for composites with Si microinclusions shown in (b). A clear redshift in the reflectance peaks is observed with an increase in the refractive index of the host medium pointing to the generation of LSPRs in the larger semiconductor microinclusions.}
\label{SiTeRAnm} 
\end{figure}

Plasmonic resonances observed in the semiconductor microinclusions can have both surface and volume modes with contributions from the magnetic or electric Mie coefficients ($a_n$ or $b_n$) or both. A key feature of the surface modes or LSPRs is the broadening and red-shifting of the scattering resonances with an increase in the particle size $d$ \cite{faucheaux}. This is seen clearly manifested to varying degrees in the Mie scattering efficiencies $Q_{\mathrm{sca}}$ for the various semiconductor microinclusion materials considered here (Figures \ref{Qscag2D}a-d, \ref{Mie}a-b, and, SI Figures 2a-b and 3). Additionally, LSPRs are also known to exhibit a red-shift with an increase in the refractive index of the host medium \cite{faucheaux, wu, katyal}. Thus, to ascertain further the nature of the plasmonic resonances observed in the spectra for the different microcomposites, we compare and contrast the optical spectra obtained using host refractive index $n_m=1.5$ ($\circ$, $\diamond$) with the results from $n_m=1.3$ ($\bullet$, $\scriptstyle\bigLozenge$). Figure \ref{SiTeRAnm}a-b shows that for composites with the larger Si and Te microinclusions there occurs a red-shift in the reflectance peaks with an increase in the refractive index of the host medium while for the smaller particles such a change is not clearly discernible. Reflectance peaks at $\lambda=4.27, 3.54, 3.27, 2.81$ and $2.67$ $\mathrm{\mu m}$ in the spectra for composites with Si microinclusions of size $d=1.68$ $\mathrm{\mu m}$ red-shift to $\lambda=4.30, 3.67, 3.32, 2.89$ and $2.71$ $\mathrm{\mu m}$ respectively with a change in the host medium refractive index from $n_m =1.3$ ($\bullet$) to $1.5$ ($\circ$) (Figure \ref{SiTeRAnm}e). For composites with Te microinclusions of size $d=1.24$ $\mathrm{\mu m}$ reflectance peaks shift from $\lambda=5.30, 4.95$ and $3.96$ ($\bullet$) to $\lambda=5.56, 5.00$ and $4.00$ $\mathrm{\mu m}$ ($\circ$) respectively for this change in the refractive index of the host medium. The notable exceptions to this red-shift occur for the broad peaks at longer wavelengths $\lambda \approx 5.8$ and $7.0$ $\mathrm{\mu m}$ for composites with Si ($d=1.68$ $\mathrm{\mu m}$) and Te ($d=1.24$ $\mathrm{\mu m}$) microparticles, respectively. The likely cause for this could either be that these peaks are associated with plasmonic resonances that are volume modes or the red-shift is masked due to the broadness of the peaks. A similar trend in the red-shifting of the peaks in the reflectance spectra associated with larger microinclusion size and a change in the refractive index of the host medium is generally observed in composites with InP, InAs, Ge, and PbS microinclusions as well (SI Figures 7a-b and 8a-b, respectively). In the case of microcomposites with PbS, InAs and Te inclusions, the weak plasmonic absorption peaks that are associated with reflectance maxima outside the main absorption band exhibit similar redshift with an increase in the host refractive index (SI Figures 7d and 8d, and, Figure \ref{SiTeRAnm}d, respectively).  Thus, there appears to be a transformation in the nature of the plasmonic resonances from volume modes for the smaller microinclusions to LSPRs for composites with the larger semiconductor microinclusions. Also, it is apparent from the results presented earlier for Mie scattering that this shift is associated with and driven by a strengthening of the magnetic modes $b_n$ characteristic of the larger particles (Figure \ref{MieABn}). For the large spherical microinclusions considered here, these resonances can thus be connected to oscillatory eddy currents generated by electromagnetic waves traveling large distances along the surface of the particles \cite{bohren1985,chylek}. 

\subsection{Reflectance efficiency of the microcomposites}
\begin{figure}[H]	
\centering
\includegraphics[width=16cm,trim={6cm, 0cm, 6cm, 0cm}]{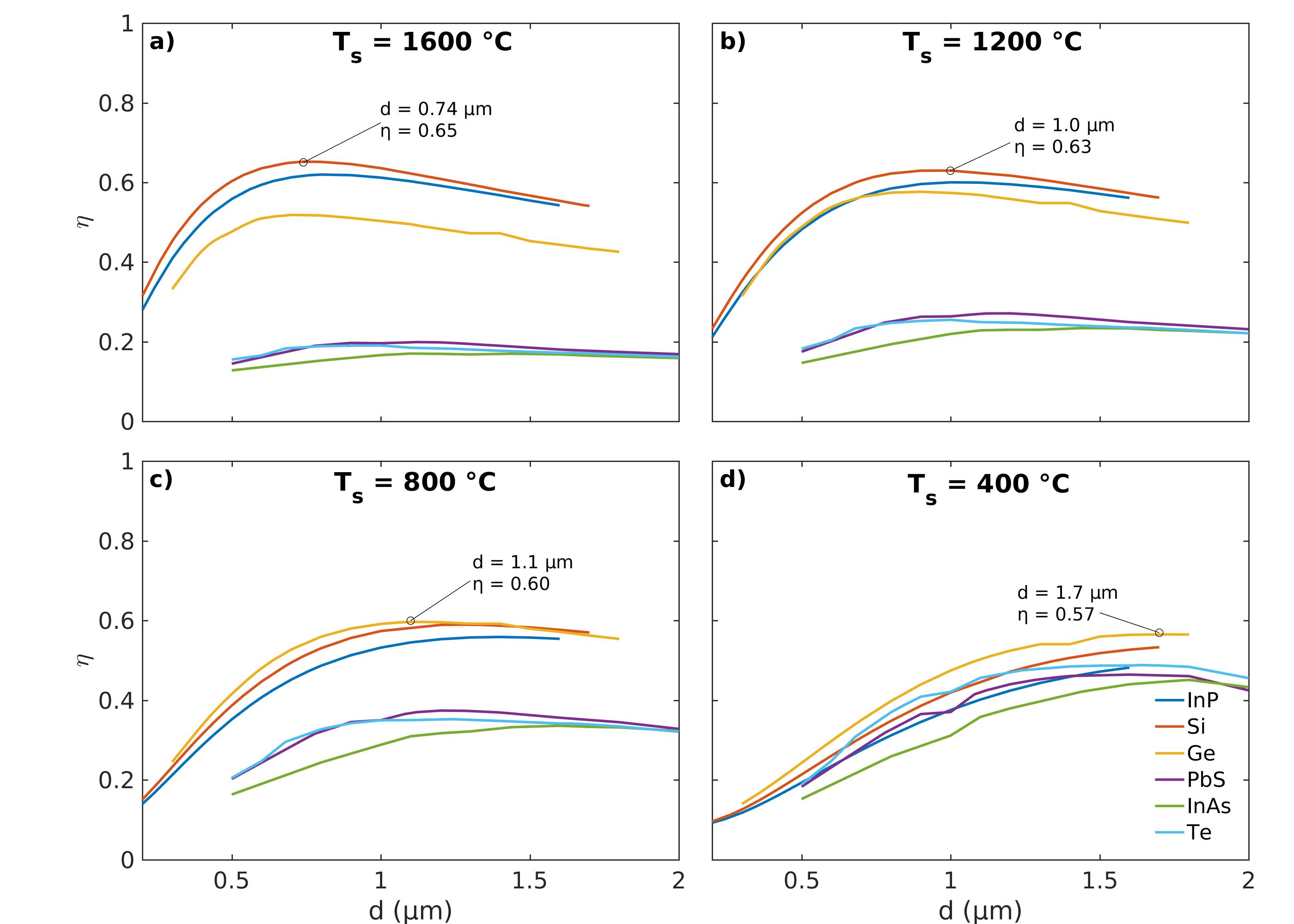}
\caption{Reflectance efficiencies $\eta$ of microcomposites with InP, Si, Ge, PbS, InAs and Te microinclusions for incident blackbody radiation from sources at temperatures in the range $400 \le T_\mathrm{s} \le 1600$ \degree C. The semiconductor microinclusions that have their bandgap wavelengths $\lambda_{\mathrm{bg}}$ close to or slightly greater than the wavelength $\lambda_{\mathrm{max}}$ of the peak spectral radiance from a blackbody source are the most effective in maximizing reflectance of the incident heat radiation.}
\label{Tregion} 
\end{figure}

To assess the effectiveness of the different microcomposite materials in preventing thermal losses through radiative transfer, the reflectance efficiency $\eta(\lambda, d)$, defined in equation \eqref{E}, is computed as a function of the size $d$ of the semiconducting microinclusions. The calculations for $\eta$ cover the entire wavelength range of interest ($\lambda=0.5$ to $10$ $\mathrm{\mu m}$) for the incident radiation from blackbody sources at temperatures $T_\mathrm{s}=1600$, 1200, 800 and 400 \degree C. Here, we note that the peak spectral radiance for a blackbody at temperatures $T_\mathrm{s}=1600, 1200$, $800$ and $400$ \degree C is obtained at $\lambda_{\mathrm{max}}=1.55, 1.97, 2.70$ and $4.31$ $\mathrm{\mu m}$ respectively. Figure \ref{Tregion} shows high values of $(0.65>\eta>0.55)$ implying reflectances of over 60\% obtained from microcomposites with an optimal size $d$ of the semiconducting microinclusions.  For the blackbody radiation from sources at temperatures $T_\mathrm{s}=1600$ and $1200$ \degree C, the highest values of efficiency $\eta = 0.65$ and $0.63$ are obtained for Si microcomposites with optimal microinclusion diameters $d=0.74$ and $1.0$ $\mathrm{\mu m}$ respectively (Figure \ref{Tregion}a-b). On the other hand, microcomposites with Ge inclusions of optimal diameters $d=1.10$ and $1.70$ $\mathrm{\mu m}$ attain the highest efficiency values of $\eta=0.60$ and $0.57$ for radiation sources characterized by temperatures $T_\mathrm{s}=800 $ and 400 \degree C respectively (Figure \ref{Tregion}c-d). These results thus show that as the wavelength $\lambda_{\mathrm{max}}$ for the peak spectral radiance increases with decreasing source temperatures, the size of the microinclusions required for obtaining peak reflectance efficiency also increases. This shift in the optimal particle diameters $d$ for obtaining maximal reflectance efficiency $\eta$ is consistent with the broadening and shifting of the peaks for $Q_{\mathrm{sca}}$ (Figures \ref{Qscag2D}a-d, \ref{Mie}a-b, and, SI Figures 2a-b and 3) and reflectance $R$  (Figures \ref{GePbSRA}a-b, \ref{SiTeRAnm}a-b and SI Figure 7a-b) towards longer wavelengths with increasing microinclusion size $d$.

Among all the semiconducting materials considered here, it is also observed that Si, Ge and InP microinclusions with larger bandgaps are the only effective inclusion materials for incident blackbody radiation from sources at temperatures in the range $400 \le T_\mathrm{s} \le 1600$ \degree C (Figure \ref{Tregion}). Furthermore, Figure \ref{Tregion} shows that the three semiconductors (PbS, InAs and Te) with the smaller bandgaps begin to significantly contribute to the reflectance efficiency $\eta$ only when their corresponding bandgap wavelength $\lambda_{\mathrm{bg}}$ becomes smaller than the wavelength ($\lambda_{\mathrm{max}}=4.31$ $\mathrm{\mu m}$) for the peak spectral radiance corresponding to the lowest source temperature $T_\mathrm{s}=400$ \degree C (Figure \ref{Tregion}d). Composites with Te microinclusions exhibit the most promising set of Mie parameters, a high $Q_{\mathrm{sca}}$ and the most negative $g_{\mathrm{min}}$ (Tables \ref{tab} and \ref{tab2}), and the highest peak in reflectance $R=0.91$ observed amongst all the semiconducting microinclusions (Figure \ref{SiTeRAnm}b). However, despite this, a high reflectance efficiency $\eta$ is not observed in Te microcomposites for any of the blackbody source temperatures $T_\mathrm{s}$ considered here because of the large bandgap wavelength $\lambda_{\mathrm{bg}}=3.75$ $\mathrm{\mu m}$ for Te (Table \ref{tab}). In contrast, InP, Si and Ge have bandgaps occurring at wavelengths $\lambda_{\mathrm{bg}}=0.92, 1.11$ and $1.85$ $\mathrm{\mu m}$ (see Table \ref{tab}) that are all smaller than the wavelengths $\lambda_{\mathrm{max}}$ for the peak spectral radiance of the blackbody source temperatures considered here. Thus, it can be inferred that semiconducting microinclusions with their bandgap wavelengths $\lambda_{\mathrm{bg}}$ close to or slightly greater than the wavelength $\lambda_{\mathrm{max}}$ of the peak spectral radiance from a blackbody source are the most effective in maximizing reflectance. This happens because close to the wavelength $\lambda_{\mathrm{bg}}$ there exist enough free charge carriers in the conduction or valence band to allow for the excitation of LSPRs that improve reflectance of the incident thermal radiation through enhanced scattering.  

\section{Conclusions}
To summarize, we have investigated the use of plasmonic resonance driven enhanced scattering from low-bandgap semiconductor microinclusions for tailoring the spectral properties of insulating composites to prevent radiative thermal losses in high temperature applications. To simulate radiative transfer in composites with semiconductor microinclusions of different materials, we have employed Monte Carlo modeling in conjunction with Mie theory. We have also compared and contrasted our results from the Monte Carlo modeling with reflectance and absorbance spectra obtained from Fresnel's equations, based on MG-EMT, that do not account for scattering from the microinclusions. Comparative results show that there is a significant enhancement in reflectance and absorbance of the incident thermal radiation due to a decrease in the average pathlength of the photons in the microcomposite layer from enhanced scattering. 

The key focus of our effort in this study has been to understand the role of the size-dependent Mie scattering ($Q_{\mathrm{sca}}$) and absorption ($Q_{\mathrm{abs}}$) efficiencies and the scattering anisotropy $g$ of microinclusions in maximizing the thermal reflectance efficiency $\eta$. Our results show that Mie coefficients of order $n \le 3$ alone contribute significantly to the Mie parameters for the spherical microinclusions. The Mie coefficients $a_n$ and $b_n$ corresponding to the electric and magnetic fields, respectively, show that the spectral features in $Q_{\mathrm{abs}}$, $Q_{\mathrm{sca}}$ and $g$ arise from the interference effects among different multipole contributions. The sharp peaks in the higher order magnetic modes for the larger microinclusions against a background of the broad dipole modes give rise to Fano resonances that generate sharp peaks in the scattering efficiency $Q_{\mathrm{sca}}$. For all semiconducting microinclusions, the first of the plasmonic resonance driven peaks in reflectance appear just outside the absorption band edge for wavelengths $\lambda \gtrsim \lambda_{\mathrm{bg}}$. The spectral features in $Q_{\mathrm{sca}}$ and $Q_{\mathrm{abs}}$ redshift and broaden with an increase in the size $d$ of the semiconducting microinclusions caused by an increase in the strength of the magnetic modes $b_n$. This redshift and broadening of spectral features is also seen in the reflectance and absorbance spectra for the different semiconducting materials used as inclusions in the insulating dielectric. For some semiconductor microinclusions (PbS, Te and InAs) a further broadbanding of the reflectance spectra is observed to be associated with absorbance peaks that are about $10-20$ times weaker as compared to the main absorption band. These absorbance peaks likely arise due to defect states within the bandgap that contribute enough charge carriers to the conduction band for plasmonic resonance driven enhanced scattering resulting in increased reflectance. A redshift in the reflectance peaks for the larger microinclusions with an increase in the refractive index of the host medium points to the transformation in the nature of the plasmonic resonances from volume modes for the smaller particles to LSPRs for the larger microinclusions. A low negative value of the scattering anisotropy $g$ lying outside the main absorption band does appear to enhance reflectance as hypothesized, but the resulting effect is not as pronounced as that from changes in $Q_{\mathrm{sca}}$ and $Q_{\mathrm{abs}}$. A high value of reflectance $R \ge 88\%$ observed in the spectra, for the different semiconducting microinclusions considered here, is in general associated with high scattering and low absorption efficiencies obtained from Mie theory. 

An increase in the volume fraction $f$ of the microinclusions or an increase in the thickness $t$ of the microcomposite lead to broadening of the reflectance at longer wavelengths that is often accompanied by an appearance of additional peaks. Results for the reflectance efficiency $\eta$ show that semiconducting microinclusions (Si, Ge and InP) with their bandgap wavelengths ($\lambda_{\mathrm{bg}}$) close to and greater than the wavelength ($\lambda_{\mathrm{max}}$) of the peak spectral radiance for incident blackbody radiation from a source at a given temperature $T_\mathrm{s}$ serves to maximize $\eta$. The highest reflectance efficiencies $0.57 \le \eta \le 0.65$, corresponding to more than $57\%$ back-reflectance, are obtained for Si and Ge microinclusions at really low volume fractions $(f=0.01)$ for incident blackbody radiation from sources at temperatures in the range $400 \le T_\mathrm{s} \le 1600$ \degree C. It is also observed that with an increase in the wavelength ($\lambda_{\mathrm{max}}$) for the peak spectral radiance a commensurate increase in the size of the semiconducting microinclusions is also required for obtaining optimal reflectance efficiency $\eta$. Thus, to fully maximize reflectance for preventing thermal losses through radiative transfer, polydispersity in the size of the microinclusions is desirable.

In conclusion, we have demonstrated that enhanced scattering due to plasmonic resonances in low-bandgap semiconductor microinclusions at really small volume fractions in an insulating dielectric can be exploited for preventing radiative thermal losses by maximizing reflectance of the incident infrared radiation in high temperature applications. Our results also suggest that the use of semiconductor microinclusions in insulating dielectrics offers a possiblity for the further enhancement and broadbanding of the reflectance spectra through the use of dopants for engineering defect states within the semiconductor bandgap that contribute to LSPRs at thermal infrared wavelengths.

\setcounter{secnumdepth}{0}
\section{Supporting Information}
Supporting Information is available from the Wiley Online Library or from the author.

\section{Acknowledgments}
The authors gratefully acknowledge funding and support from the Academy of Finland, Center of Excellence Programme (2015-2017), Project No. 284621; the Aalto Energy Efficiency Research Program EXPECTS; and, the Aalto Science-IT project. VT also gratefully acknowledges useful discussions with M.Eng. Dimitrios Tzarouchis on Mie scattering.

\providecommand*{\mcitethebibliography}{\thebibliography}
\csname @ifundefined\endcsname{endmcitethebibliography}
{\let\endmcitethebibliography\endthebibliography}{}

\end{document}


\Large
\title{\textbf{Supporting Information: Plasmonically Enhanced Reflectance of Heat Radiation from Low-Bandgap Semiconductor Microinclusions}}
\large
\author[1*]{Janika Tang} 
\author[1*]{Vaibhav Thakore} 
 \author[1,2,3]{Tapio Ala-Nissila}
\affil[1]{COMP CoE at the Department of Applied Physics, Aalto University School of Science, FIN-00076 Aalto, Espoo, Finland}
\affil[2]{Department of Physics, Brown University, Providence, Rhode Island 02912-1843, USA}
\affil[3]{Department of Mathematical Sciences and Department of Physics, Loughborough University, Loughborough LE11 3TU, UK}
\maketitle
\text{*Email: vthakore@knights.ucf.edu} (Corresponding author)\\
\text{*Email: janika.tang@aalto.fi}  (Corresponding author)\\
\normalsize 
\doublespacing

\setcounter{secnumdepth}{3}

\section{Modification of the Monte Carlo method for modeling thermal radiation transport}

The Monte Carlo method developed by Wang et al. \cite{mcml} for modeling radiation transport in multilayered turbid media is modified and adapted to compute spectral transmittance, reflectance and absorbance for a free-standing layer of insulating dielectric composite with low-bandgap semiconducting microinclusions. The method by Wang et al. \cite{mcml} for the  computation of the specular reflectance considers the first layer of the multilayer system to be non-absorbing. However, this is not valid for the case of a freely suspended absorbing composite layer considered in our work and thus the specular reflectance $R_{\mathrm{sp}}$ is modified to

\begin{equation}
R_{\mathrm{sp}}=\frac{(n_0-n)^2+\kappa^2}{(n_0+n)^2+\kappa^2},
\end{equation}

where $n$ and $\kappa$ are the real and imaginary parts of the refractive index and $n_0$ is the refractive index of the non-absorbing ambient medium.

To validate the modified Monte Carlo model, we performed simulations for dielectric composites with titanium dioxide $(\mathrm{TiO}_2)$ and vanadium dioxide $(\mathrm{VO}_2)$ nanoparticle inclusions of various sizes. The corresponding results were then compared against the results obtained using the Fresnel equations and the four-flux method by Laaksonen et al. \cite{laaksonen2014}. The optical constants for $\mathrm{TiO}_2$ and the semiconducting and metallic phases of thermochromic $\mathrm{VO}_2$ were obtained from the references \cite{palik, mlyuka} respectively. The nanoparticle inclusions were assumed to be embedded in a host medium of refractive index $n_m = 1.5$. The volume fraction of the nanoinclusions used in the simulations and the composite layer thickness were specified as $f = 0.01$ and $t=10$ $\mathrm{\mu m}$ respectively. A grid resolution of $dz = 0.1$ $\mathrm{\mu m}$ and $dr = 5$ $\mathrm{\mu m}$ was used for the radial $\hat{z}$ and axial $\hat{r}$ directions respectively (main text, see Figure 1a). The total number of grid elements in the $\hat{r}$-direction was set to $N_r=100$ while the number of grid elements $N_z$ in the $\hat{z}$-direction was determined by the thickness of the microcomposite layer. The angular dependence of the reflectance and transmittance spectra on the photon-exiting direction $\hat{\alpha}$ was ignored. Each simulation was carried out using $10^7$ photons. Similar to the computation of the optical parameters by Laaksonen \textit{et al.} \cite{laaksonen2014}, the scattering $\mu_{\mathrm{sca}}$ and absorption $\mu_{\mathrm{abs}}$ coefficients, and, the scattering anisotropy $g$ were computed using Mie theory for use with the Monte Carlo method (main text, see Figure 1b). As in Laaksonen \textit{et al.} \cite{laaksonen2014}, unlike the study presented here, the optical parameters employed in the Fresnel equations were computed using equations (3) and (7) in the main text based on the MG-EMT for validation.

\begin{figure}[H]	
\centering
\includegraphics[width=13.5cm, trim={7cm, 0cm, 7cm, 0cm}]{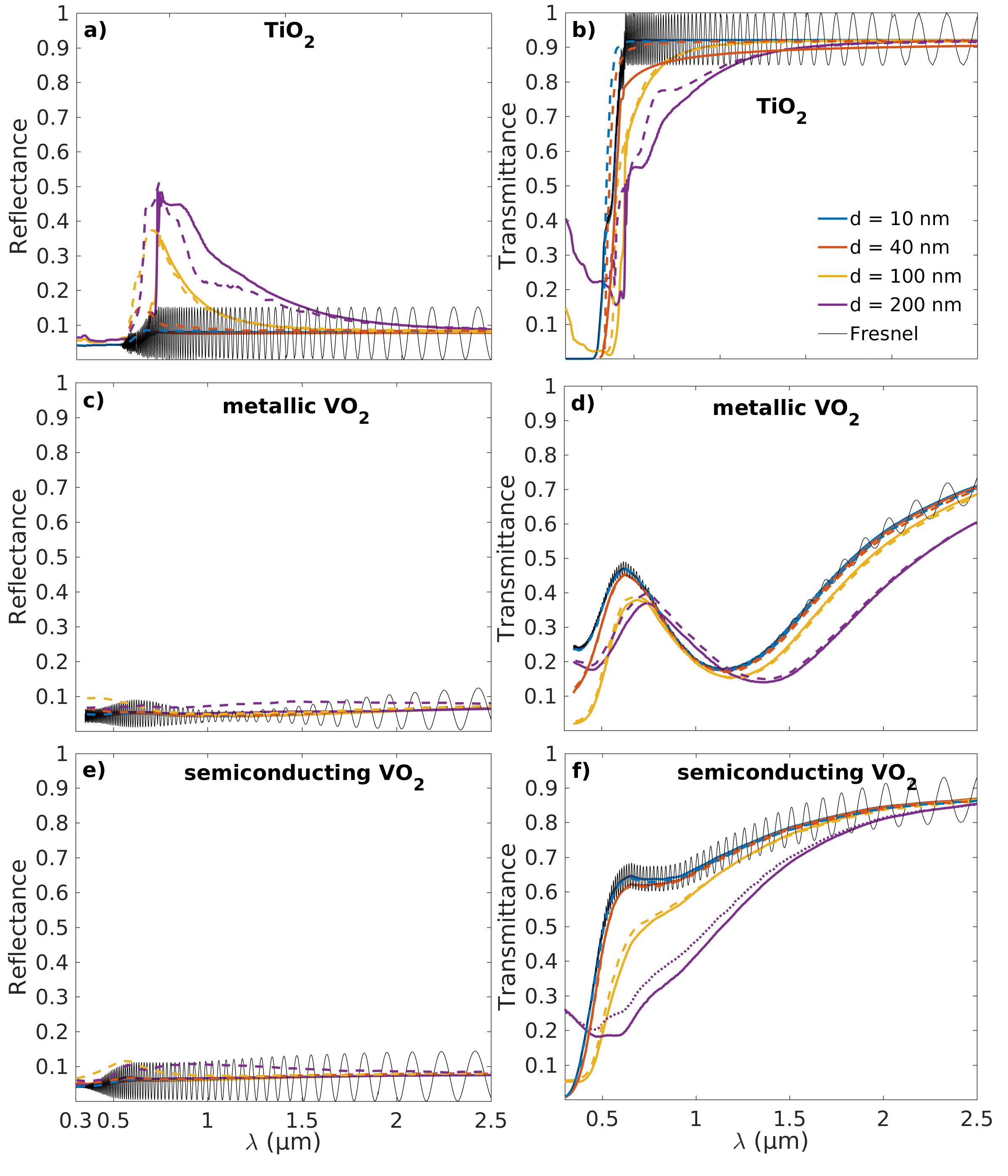}
\caption{Spectral transmittance and reflectance for (a-b) $\mathrm{TiO}_2$, (c-d) metallic $\mathrm{VO}_2$, and (e-f) semiconducting $\mathrm{VO}_2$ nanoparticles with volume fraction $f=0.01$ embedded in a dielectric host medium of refractive index $n=1.5$ and thickness $t=10$ $\mathrm{\mu m}$ calculated using the Monte Carlo method (solid lines), the four-flux method (dashed lines) and the Fresnel equations (black lines).}
\label{Verification} 
\end{figure}

SI Figure \ref{Verification}a-f shows comparisons of the transmittance and reflectance spectra for the $\mathrm{TiO}_2$ and $\mathrm{VO}_2$ nanocomposites computed using the Fresnel equations \cite{heavens}, the modified Monte Carlo and the four-flux methods \cite{laaksonen2014}. A comparison between the computed spectra from the four-flux and the Monte Carlo methods shows that the agreement with the results from the Fresnel equations is better for the Monte Carlo model at smaller wavelengths in the regime of low scattering (SI Figure \ref{Verification}). Both the four-flux and the Monte Carlo methods capture the average spectral behavior quite accurately for the smaller nanoinclusions of size $d=\SI{10}{\nano\meter}$ at longer wavelengths wherein interference effects dominate the optical spectra obtained from the Fresnel equations. It must be noted that unlike the Fresnel equations both the four-flux and the Monte Carlo models of radiative transfer do not account for the interference effects. The deviations from the results obtained using Fresnel equations for the four-flux and the Monte Carlo methods become more significant due to increased scattering effects with an increase in the size of the nanoinclusions. There are also some differences between the results from the four-flux and the Monte Carlo methods. These differences in transmittance and reflectance spectra are more pronounced in the case of composites with $\mathrm{TiO}_2$ nanoinclusions (SI Figure \ref{Verification}a-b) that exhibit enhanced scattering while the agreement between the two methods is much better for the metallic and semiconducting forms of the thermochromic $\mathrm{VO}_2$ nanocomposites characterized by lower scattering (SI Figure \ref{Verification}c-f). This difference in optical spectra obtained from the four-flux and the Monte Carlo methods is observed because of the assumptions made in the four-flux method with regard to the value of the average path-length parameter and the ratio of the forward scattering radiation to that of the collimated light \cite{laaksonen2014}. The Monte Carlo method, on the other hand, does not make any such assumptions and is known to be more accurate than the four-flux-method \cite{maheu}.

\section{Mie scattering from semiconductor microinclusions}

\begin{figure}[H]	
\centering
\includegraphics[width=15cm, trim={6cm, 0cm, 6cm, 0cm}]{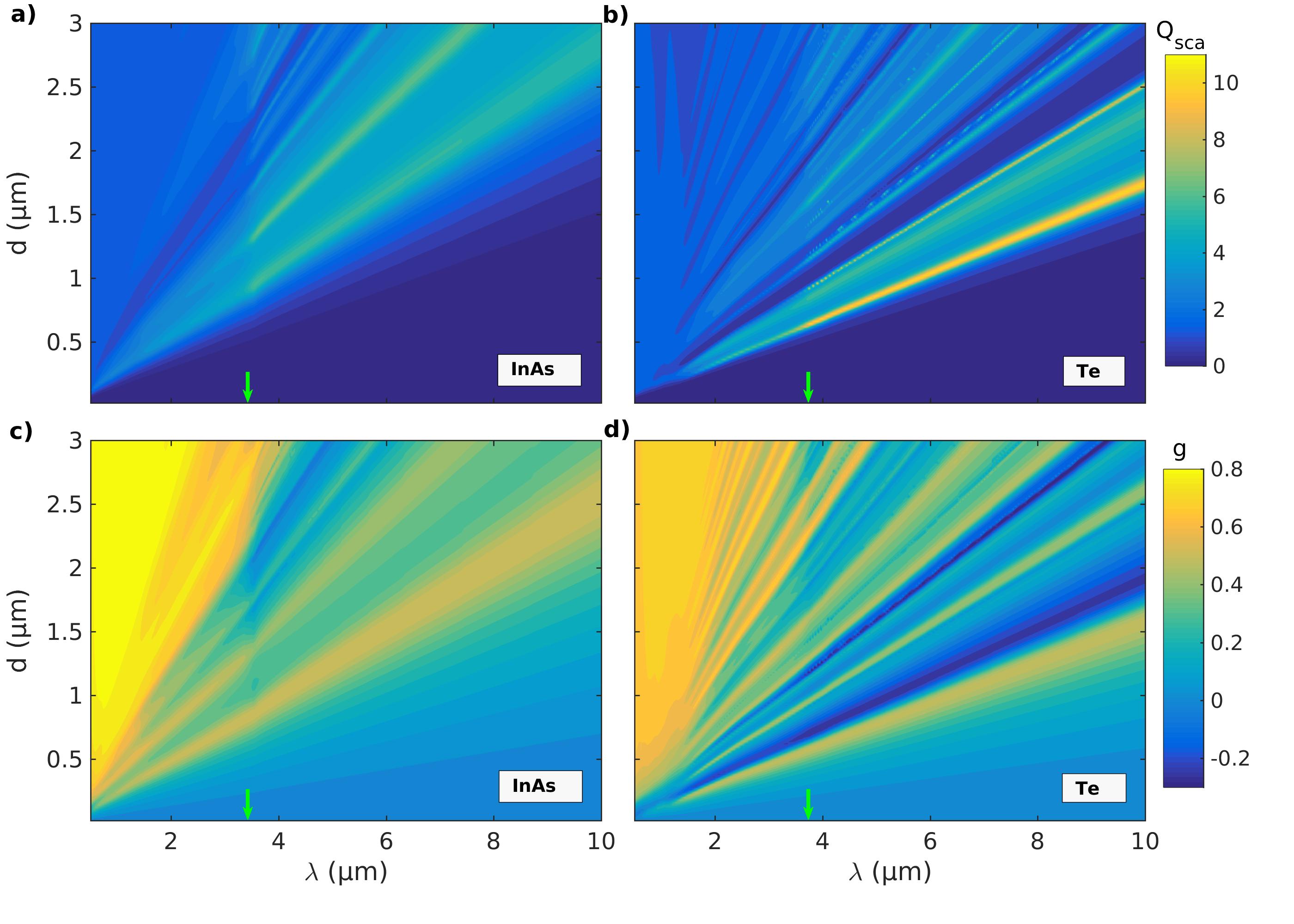}
\caption{(a-b) Scattering efficiency $Q_{\mathrm{sca}}$, and, (c-d) anisotropy factor $g$ as a function of the wavelength $\lambda$ of the incident thermal radiation and the diameter $d$ of spherical InAs and Te microinclusions, respectively. Similar to the plots of $Q_{\mathrm{sca}}$ and $g$ for InP, Si, Ge and PbS microinclusions (main text, Figure 4), the bandgap wavelengths $\lambda_{\mathrm{bg}}$ (indicated by vertical green arrows) for the semiconductor materials here denote a transition from low to high $Q_{\mathrm{sca}}$ and strongly forward ($+g$) to mixed scattering regimes for the microinclusions with increasing $\lambda$.} 
\label{Qscag2D} 
\end{figure}

\begin{figure}[H]	
\centering
\includegraphics[width=17cm, trim={6cm, 0cm, 6cm, 0cm}]{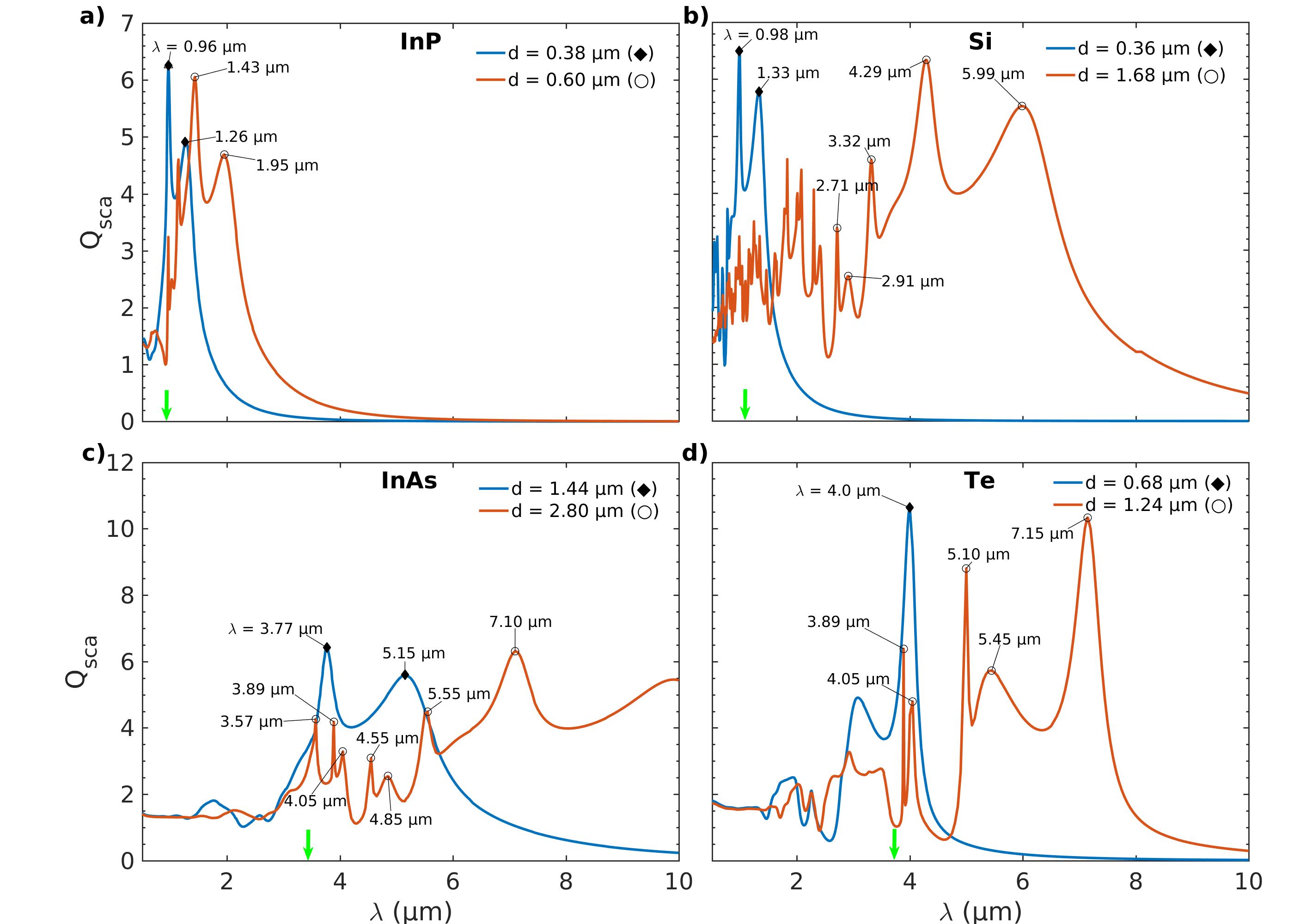}
\caption{Mie scattering efficiencies $Q_{\mathrm{sca}}$ for (a) InP, (b) Si, (c) InAs, and, (d) Te microinclusions of sizes $d$ corresponding to $Q^{\mathrm{max}}_{\mathrm{sca}}$ and $g_{\mathrm{min}}$, respectively (main text, Figure 3 and Tables 1 and 2). The vertical green arrows indicate the bandgap wavelengths $\lambda_{\mathrm{bg}}$. Sharp Fano resonances in $Q_{\mathrm{sca}}$ are observed for the composites with Si, InAs and Te microinclusions with an increase in the particle size.}
\label{QscaLine} 
\end{figure}

\begin{figure}[H]	
\centering
\includegraphics[width=17cm, trim={6cm, 0cm, 6cm, 0cm}]{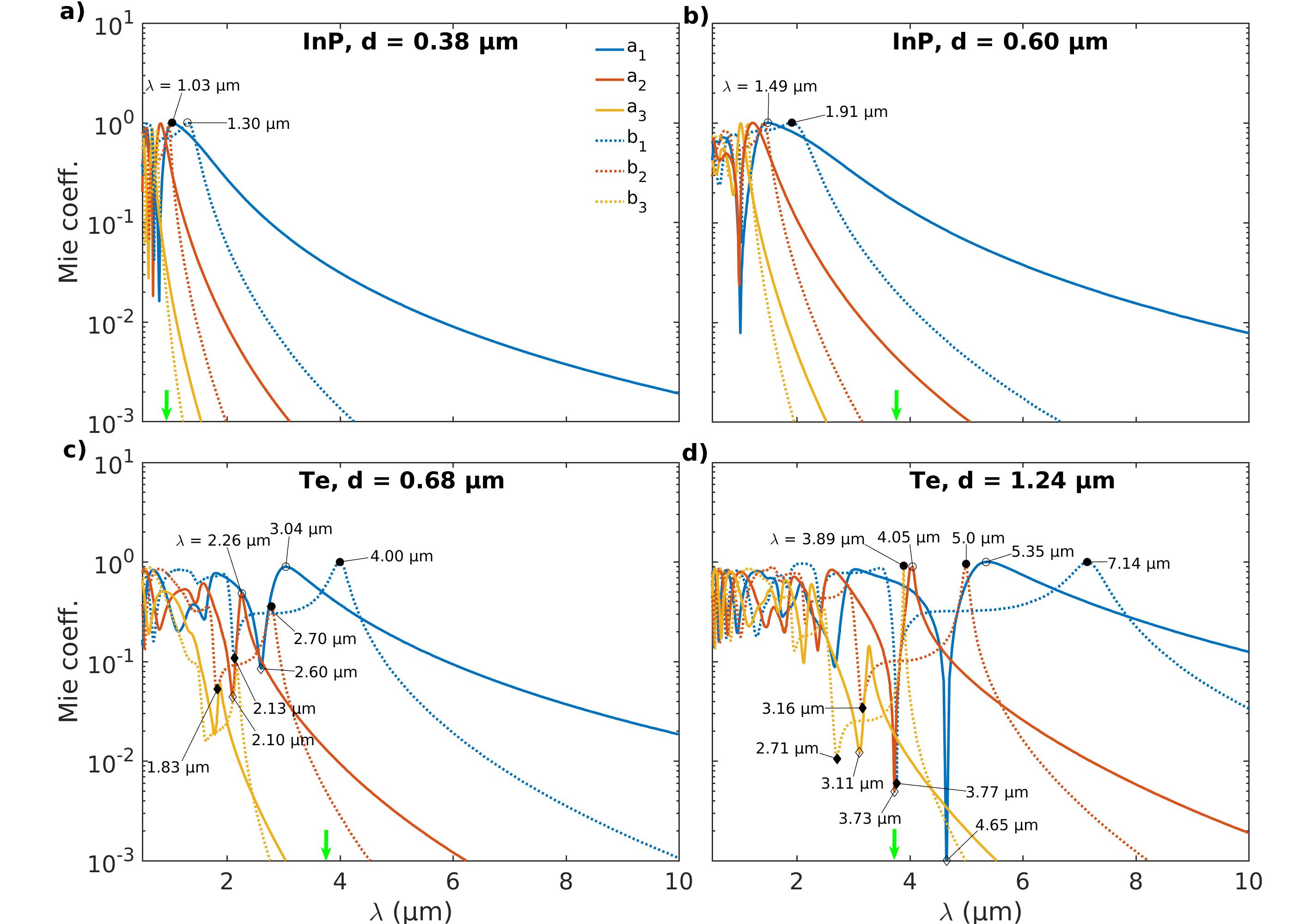}
\caption{Mie coefficients $a_{n}$ (solid-line) and $b_{n}$ (dashed-line) as a function of wavelength $\lambda$ for spherical microinclusions of (a, b) InP, and, (c,d) Te for particle diameters $d$ corresponding to $Q^{\mathrm{max}}_{\mathrm{sca}}$ and $g_{\mathrm{min}}$, respectively (main text Figure 2 and Tables 1 and 2). The vertical green arrows indicate the bandgap wavelengths $\lambda_{\mathrm{bg}}$. Similar to the results reported for Si, Ge, PbS and InAs microinclusions (main text, Figure 5), an increase in the microinclusion size $d$ is accompanied by a strengthening of the magnetic modes $b_n$.}
\label{TeInPMieAnBn} 
\end{figure}

\begin{figure}[H]	
\centering
\includegraphics[width=13cm,trim={8cm, 0cm, 8cm, 0cm}]{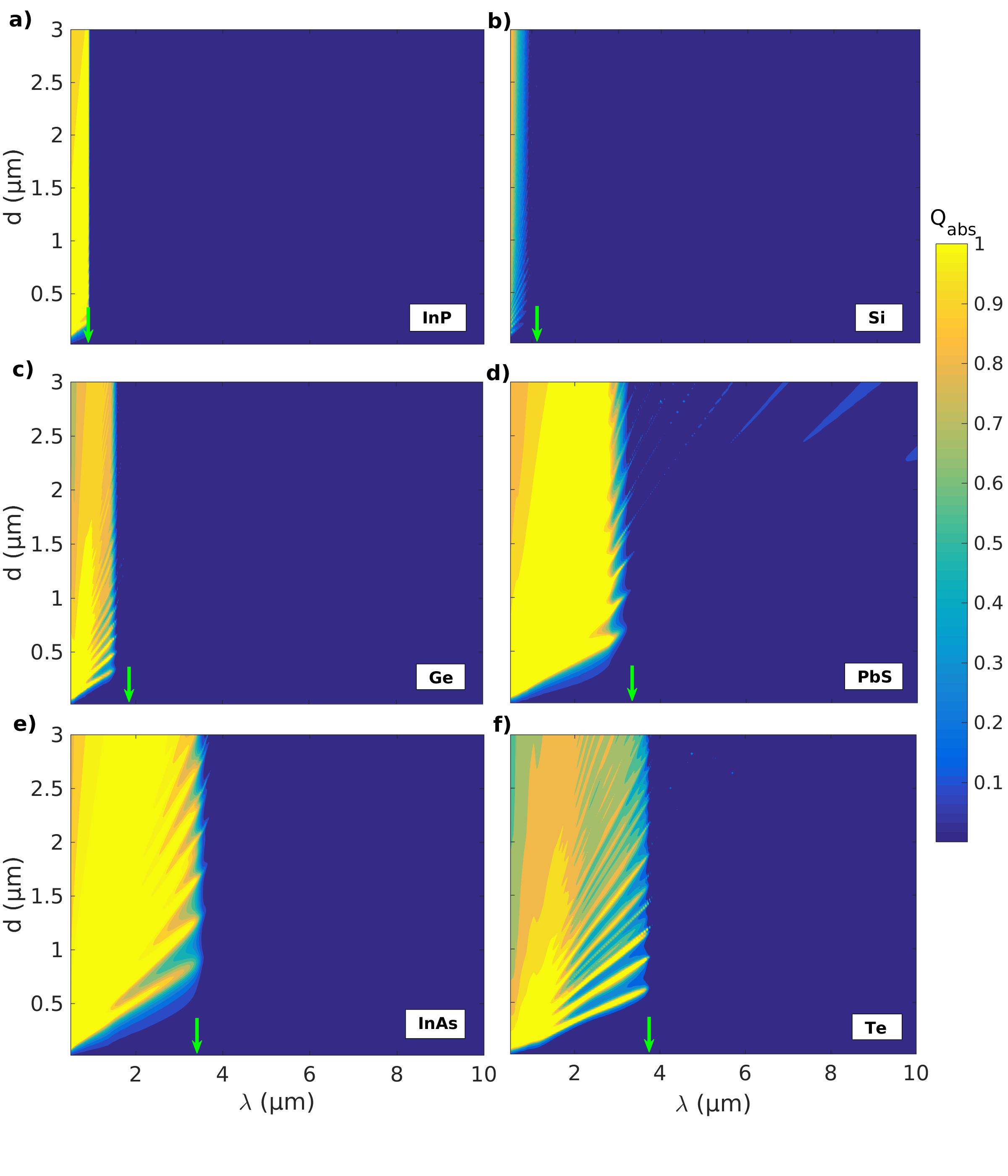}
\caption{Absorption efficiency $Q_{\mathrm{abs}}$ as a function of wavelength $\lambda$ and diameter $d$ of spherical (a) InP, (b) Si, (c) Ge, (d) PbS, (e) InAs, and, (f) Te microinclusions. The vertical green arrows on the x-axis indicate the bandgap wavelengths $\lambda_{\mathrm{bg}}$ marking the onset of the absorption edge.}
\label{Qabs2D} 
\end{figure}

\begin{figure}[H]	
\centering
\includegraphics[width=16cm,trim={8cm, 0cm, 8cm, 0cm}]{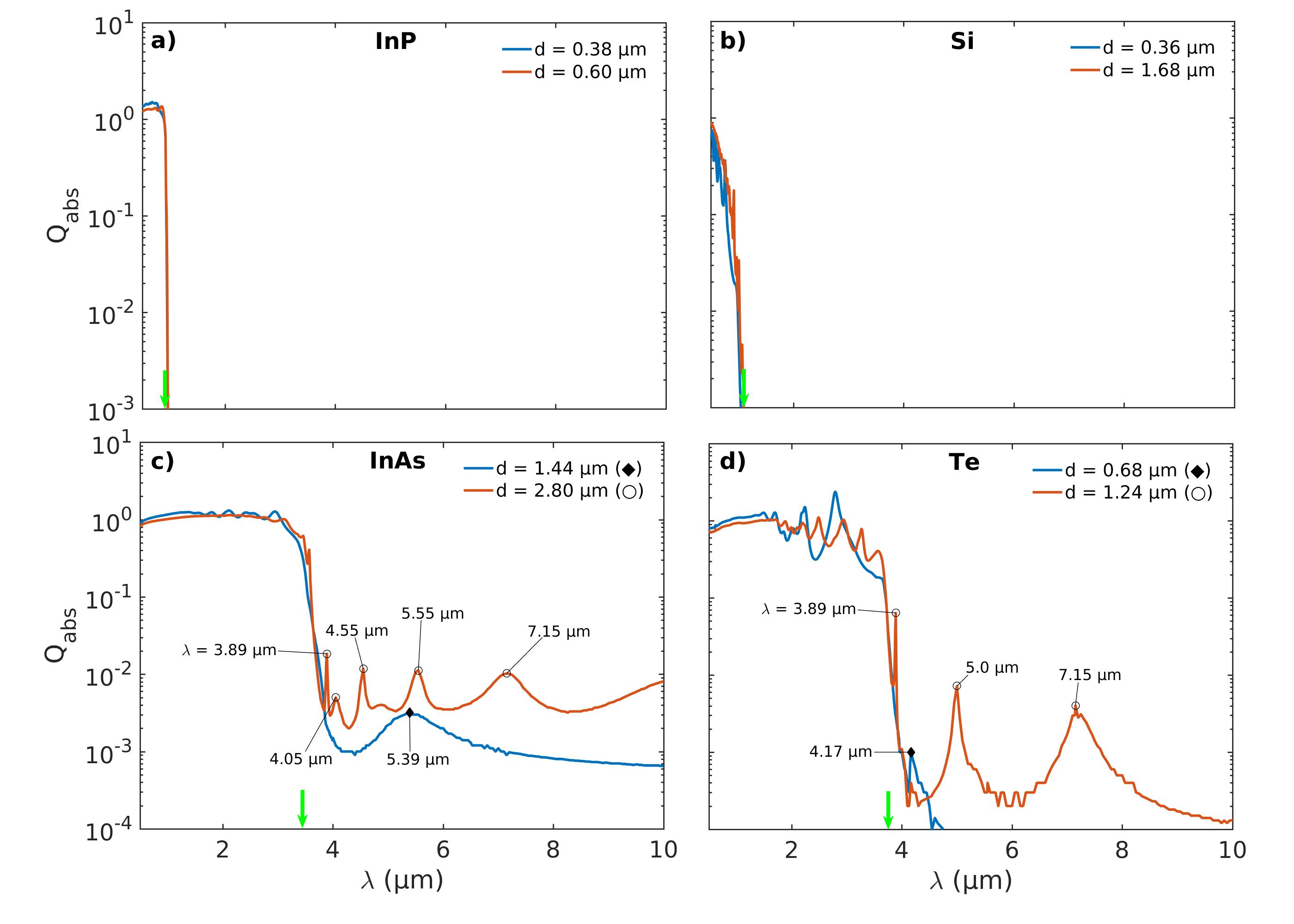}
\caption{Absorption efficiency $Q_{\mathrm{abs}}$ for various sizes of (a) InP, (b) Si, (c) InAs and (d) Te microinclusions of sizes $d$ corresponding to $Q^{\mathrm{max}}_{\mathrm{sca}}$ and $g_{\mathrm{min}}$, respectively (main text, Figure 2 and Tables 1 and 2). The vertical green arrows indicate the bandgap wavelengths $\lambda_{\mathrm{bg}}$. Contributions to $Q_{\mathrm{abs}}$ due to weak absorption from the defect states in composites with InAs and Te microinclusions serve to significantly broadband the reflectance of the incident thermal radiation.}
\label{QabsLine} 
\end{figure}

SI Figures \ref{Qscag2D}a-d show the scattering efficiency $Q_{\mathrm{sca}}$ and the anisotropy factor $g$ for Te and InAs microinclusions of diameters varying between $d=0.02$ to $3$ $\mathrm{\mu m}$ in the wavelength range $\lambda=0.5$ to $10$ $\mathrm{\mu m}$ for the incident thermal radiation. The scattering anisotropy $g$ is observed to be strongly forward-scattering after the absorption edge for sizes of microinclusions that are comparable to the wavelength of the incident thermal radiation (SI Figure \ref{Qscag2D}c-d). Similar to the PbS microinclusions (main text, Figures 3d, and, 4b, d), free charge carriers created in the conduction or valence bands at $\lambda \gtrsim \lambda_{\mathrm{bg}}$ (SI Figure \ref{Qabs2D}e-f) and due to weak absorption peaks (SI Figure \ref{QabsLine}c-d) outside of the main absorption band at longer wavelengths result in plasmonic resonances that give rise to large peaks in $Q_{\mathrm{sca}}$ for Te and InAs microinclusions (SI Figures \ref{Qscag2D}a-b and \ref{QscaLine}c-d, respectively). As pointed out in the main text, the features in $Q_{\mathrm{sca}}$ and $g$, such as maxima and minima, are seen to redshift and broaden to varying degrees with an increase in the size $d$ of the microinclusions for all materials studied here (main text, Figure 3 and SI Figures \ref{Qscag2D} and \ref{QscaLine}). Also, it is clearly seen that $Q_{\mathrm{sca}}$ in SI Figure \ref{Qscag2D}a-b obtains lower values for the wavelengths $\lambda < \lambda_{\mathrm{bg}}$ for both InAs and Te. The reduction in $Q_{\mathrm{sca}}$ for InAs and Te microinclusions follows an increase in the absorption as seen in SI Figure \ref{Qabs2D}e-f. 

SI Figure \ref{TeInPMieAnBn} presents the Mie coefficients of modes $n\le3$ for InP and Te microinclusions of size $d_{Q_{\mathrm{sca}}^{\mathrm{max}}}$ and $d_{g_{\mathrm{min}}}$ that correspond to the highest and the lowest values of the scattering efficiency $Q_{\mathrm{sca}}^{\mathrm{max}}$ and the anisotropy factor $g_{min}$ respectively (Tables 1 and 2, main text). It can be seen that the sharp peaks in the magnetic and electric modes give rise to peaks in scattering efficiencies $Q_{\mathrm{sca}}$. For example, the peaks in Mie coefficients at $\lambda = 7.14, 5.35, 5.0, 4.05$ and $3.89$ $\mathrm{\mu m}$ (SI Figure \ref{TeInPMieAnBn}b) for Te microinclusions of size $d=1.24$ $\mathrm{\mu m}$ ($\circ$) contribute to resonances in $Q_{\mathrm{sca}}$ at $7.15, 5.45, 5.0, 4.05$ and $3.89$ $\mathrm{\mu m}$ (SI Figure \ref{QscaLine}d) respectively.

\section{Spectral reflectance of microcomposites}
\begin{figure}[H]	
\centering
\includegraphics[width=17cm,trim={6cm, 0cm, 6cm, 0cm}]{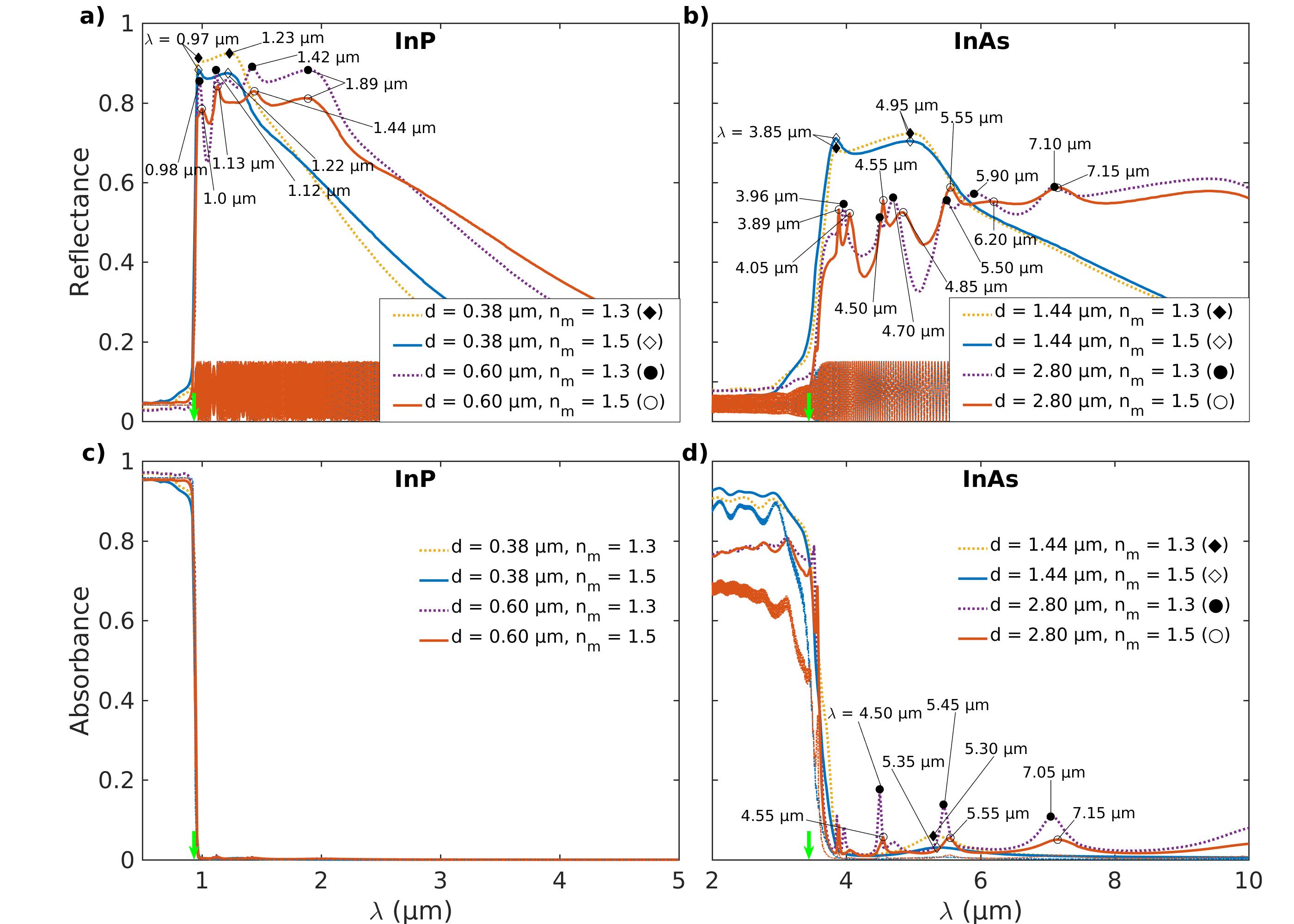}
\caption{The spectral reflectance and absorbance of microcomposites with (a, c) InP, and (b, d) InAs spherical inclusions of diameter $d$ and volume fraction $f=0.01$ embedded in a dielectric host medium of refractive indices $n_m=1.5$ and $1.3$ and thickness $t=200$ $\mathrm{\mu m}$. The solid lines correspond to results from Monte Carlo simulations for the case of host refractive index $n_m=1.5$ while the thin dotted lines of the same color denote results from Fresnel equations. The vertical green arrows on the x-axis indicate the bandgap wavelengths $\lambda_{\mathrm{bg}}$. The broadbanding of the reflectance peaks in microcomposites with InAs inclusions correlates well with the weak peaks in absorbance that occur away from the main absorption band at longer wavelengths.}
\label{InPInAsRA} 
\end{figure}

Scattering and absorption from the low-bandgap semiconducting microinclusions significantly increases reflectance and absorbance as can be seen from a comparison of the spectra obtained using the Monte Carlo method and the Fresnel equations in SI Figure \ref{InPInAsRA}. Again, the difference between the two methods emphasizes the huge impact a small volume fraction of particle inclusions has on the infrared spectra of the microcomposites with semiconductor microinclusions. High and broad maxima in reflectance spectra are observed for both InP and InAs microcomposites especially for the larger microinclusions (SI Figure \ref{InPInAsRA}a-b). 

For composites with InAs microinclusions, similar to composites with PbS particles (main text, Figure 6b), the peaks in reflectance (SI Figure \ref{InPInAsRA}b) that occur beyond the absorption band edge at longer wavelengths ($n_m = 1.5$ ($\circ$, $\diamond$); $\lambda=7.15, 5.55, 4.85$ and $4.55$ $\mathrm{\mu m}$ for $d = 2.80$ $\mathrm{\mu m}$ ($\circ$); and, $\lambda=4.95$ and $3.85$ $\mathrm{\mu m}$ for $d=1.44$ $\mathrm{\mu m}$ ($\diamond$)) correlate well with the resonances in $Q_{\mathrm{sca}}$ ($\lambda=7.10, 5.55, 4,85$ and $4.55$ $\mathrm{\mu m}$ for $d=2.80$  $\mathrm{\mu m}$($\circ$), and, $\lambda=5.15$ and $3.77$ $\mathrm{\mu m}$ for $d=1.44$ $\mathrm{\mu m}$ ($\scriptstyle\bigLozenge$), SI Figure \ref{QscaLine}c). These scattering resonances are observed to arise due to enhanced scattering from the excitation of plasmonic resonances ($\circ$, $\bullet$) that are identified in the plots for Mie coefficients (main text, Figure 5g-h and SI Figure \ref{TeInPMieAnBn}c-d). On the other hand, similar to composites with Ge microparticles (main text, Figure 5c-d), the InP microinclusions do not exhibit any plasmonic resonances in the Mie coefficients $a_n$ ($\circ$) and $b_n$ ($\bullet$) away from the main absorption band (SI Figure \ref{TeInPMieAnBn}a-b). This behavior is consistent with the absence of absorption away from the main absorption band in InP microinclusions (SI Figures \ref{QabsLine}a and \ref{InPInAsRA}c). Thus, the reflectance spectra for microcomposites with Ge (main, text Figure 6a) and InP inclusions (SI Figure \ref{InPInAsRA}a) is marked by an absence of the broadbanding that is observed for composites with PbS (main text, Figure 6b), InAs (SI Figure 7b) and Te microinclusions (main text, Figure 7a). However, the peaks in reflectance ($n_m = 1.5$ ($\circ$, $\diamond$); $\lambda=1.89, 1.44, 1.13$ and $1.0$  $\mathrm{\mu m}$ for $d=0.60$ $\mathrm{\mu m}$ ($\circ$), and, $\lambda=1.22$ and $0.97$ $\mathrm{\mu m}$ for $d=0.38$ $\mathrm{\mu m}$ ($\diamond$), SI Figure \ref{InPInAsRA}a) that occur for microcomposites with InP inclusions correlate well with peaks in $Q_{\mathrm{sca}}$ ($\lambda=1.95, 1.43, 1.14$ and $0.96$ $\mathrm{\mu m}$ for $d=0.60$ $\mathrm{\mu m}$ ($\circ$), and $\lambda=1.26$ and $0.96$  $\mathrm{\mu m}$ for $d=0.38$ $\mathrm{\mu m}$ ($\scriptstyle\bigLozenge$), SI Figure \ref{QscaLine}a) that arise from the resonances ($\circ$, $\bullet$) observed in the Mie coefficients (SI Figure \ref{TeInPMieAnBn}a-b). Again, similar to other microcomposites, a redshifting and broadening of reflectance peaks is also observed for composites with InP and InAs microinclusions with an increase in the particle size $d$ (SI Figure \ref{InPInAsRA}a-b). 

\section{Nature of plasmonic resonances}
\begin{figure}[H]	
\centering
\includegraphics[width=17cm,trim={6cm, 0cm, 6cm, 0cm}]{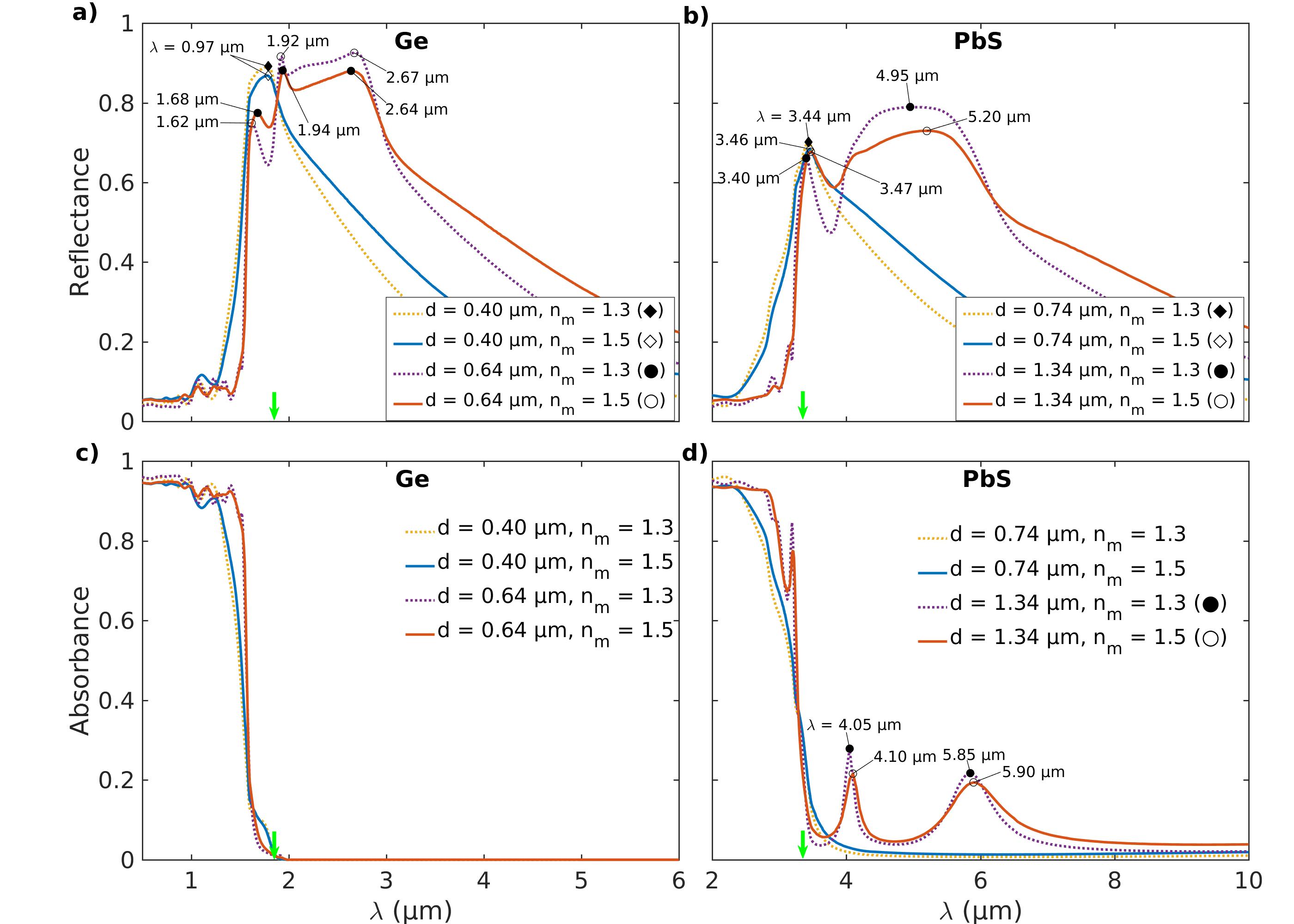}
\caption{The spectral reflectance and absorbance of microcomposites with (a, c) Ge and (b, d) PbS spherical inclusions of diameter $d$ and volume fraction $f=0.01$ embedded in dielectric media of refractive indices $n_m=1.5$  (solid-lines) and $1.3$ (dotted-lines). The vertical green arrows on the x-axis indicate the bandgap wavelengths $\lambda_{\mathrm{bg}}$. A red-shift in the reflectance peaks denotes the surface-localized nature of plasmonic resonances.}
\label{RAnm} 
\end{figure}

SI Figures \ref{InPInAsRA}a-b and \ref{RAnm} show a comparison of the reflectance spectra for composites with InP, InAs and Ge, PbS microinclusions with an increase in the refractive index of the host medium from $n_m=1.3$ ($\bullet$, $\scriptstyle\bigLozenge$) to $1.5$ ($\circ$, $\diamond$). Results show that for the smaller microinclusions there occurs no discernible redshift in the reflectance peaks with an increase in the host-medium refractive index. For example, the position of the reflectance peaks at $\lambda=0.97$ and $1.23$ $\mathrm{\mu m}$ remains unchanged for microcomposites with InP inclusions of size $d=0.38$ $\mathrm{\mu m}$ ($\scriptstyle\bigLozenge$, $\diamond$) with an increase in the host refractive index (Figure \ref{InPInAsRA}a). Similarly, in the case of composites with InAs particles of size $d=1.44$ $\mathrm{\mu m}$ ($\scriptstyle\bigLozenge$, $\diamond$) the reflectance peaks at $\lambda=3.85$ and $4.95$ $\mathrm{\mu m}$ do not exhibit any redshift (Figure \ref{InPInAsRA}b). However, for the larger microinclusions ($\bullet$, $\circ$) a distinct red shift of the reflectance peaks is seen in all microcomposites except perhaps for the broadest of peaks that occur at longer wavelengths (main text, Figure 7a-b and SI Figures \ref{InPInAsRA}a-b, \ref{RAnm}a-b). Again taking microcomposites with InP particles of size $d=0.60$ $\mathrm{\mu m}$ ($\bullet$, $\circ$) as an example it is observed that the reflectance peaks at $\lambda=1.42, 1.12$ and $0.98$ $\mathrm{\mu m}$ redshift to $\lambda=1.44, 1.13,$ and $1.0$ $\mathrm{\mu m}$ (Figure \ref{InPInAsRA}a). For the broadest reflectance peak at $\lambda \approx 1.9$ $\mathrm{\mu m}$, however, it is difficult to make out the redshift (Figure \ref{InPInAsRA}a). Thus, these results point to a transformation in the nature of plasmonic resonances from volume modes to localized surface modes driven by a strengthening of the magnetic Mie modes (main text, Figure 5 and SI Figure 4) as pointed out in the main text.    

\providecommand*{\mcitethebibliography}{\thebibliography}
\csname @ifundefined\endcsname{endmcitethebibliography}
{\let\endmcitethebibliography\endthebibliography}{}